\documentclass[journal]{vgtc}                     % final (journal style)
\onlineid{1632}

\vgtccategory{Research}

\usepackage[normalem]{ulem}
\usepackage{balance}

\newcommand\myparagraph[1]{ \vspace{4pt} \noindent \textbf{#1.}}

\newcommand{\claudio}[1]{}
\newcommand{\vitoria}[1]{}
\newcommand{\harish}[1]{}
\newcommand{\felipe}[1]{}

\newcommand{\hide}[1]{}
\newcommand{\hl}[1]{#1}

\newcommand{\add}[1]{#1}

\newcommand{\hlalgo}[1]{{\color{blue} #1}}

\newcommand{\C}{\mathbb{C}\xspace}

\newcommand{\denselist}{\itemsep 0pt\parsep=1pt\partopsep 0pt}

\title{TopoMap++: A faster and more space efficient technique to compute projections with topological guarantees}

\author{%
    \authororcid{Vitoria Guardieiro}{0000-0003-1956-5418}, \authororcid{Felipe Inagaki de Oliveira}{0009-0009-4541-2475}, \authororcid{Harish Doraiswamy}{0000-0003-2995-250X}, \authororcid{Luis Gustavo Nonato}{0000-0002-8514-8033}, \authororcid{Claudio Silva}{0000-0003-2452-2295}
}

\authorfooter{
  \item
  	Vitoria Guardieiro, Felipe Inagaki de Oliveira, and Claudio Silva are with New York University.
  	E-mail: {vitoria.guardieiro, felipe.oliveira, csilva}@nyu.edu.
  \item
  	Luis Gustavo Nonato is with ICMC-USP, São Carlos, Brazil.
  	E-mail: gnonato@icmc.usp.br.

  \item Harish Doraiswamy is with Microsoft Research India.
  	E-mail: harish.doraiswamy@microsoft.com.
}

\abstract{
High-dimensional data, characterized by many features, can be difficult to visualize effectively. Dimensionality reduction techniques, such as PCA, UMAP, and t-SNE, address this challenge by projecting the data into a lower-dimensional space while preserving important relationships. TopoMap is another technique that excels at preserving the underlying structure of the data, leading to interpretable visualizations. In particular, TopoMap maps the high-dimensional data into a visual space, guaranteeing that the 0-dimensional persistence diagram of the Rips filtration of the visual space matches the one from the high-dimensional data. However, the original TopoMap algorithm can be slow and its layout can be too sparse for large and complex datasets. In this paper, we propose three improvements to TopoMap: 1) a more space-efficient layout, 2) a significantly faster implementation, and 3) a novel TreeMap-based representation that makes use of the topological hierarchy to aid the exploration of the projections. 
These advancements make TopoMap, now referred to as TopoMap++, a more powerful tool for visualizing high-dimensional data which we demonstrate through different use case scenarios.
}

\keywords{Topological data analysis, Computational topology, High-dimensional data, Projection.}

\teaser{
\centering
\includegraphics[width=\linewidth]{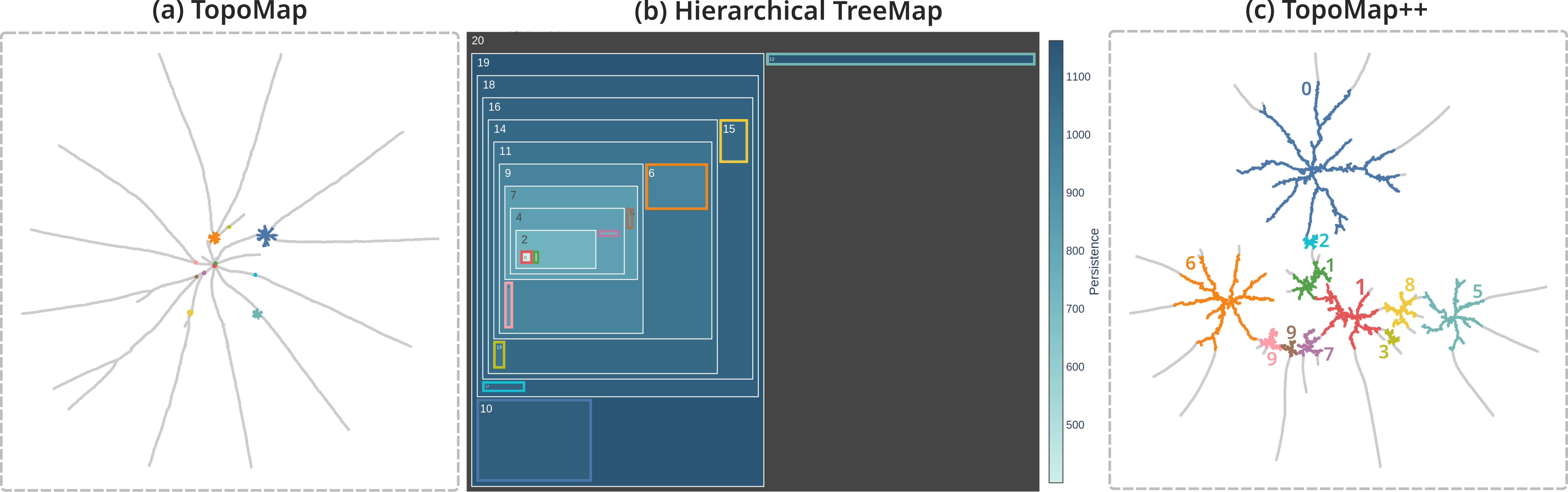}
\caption{Representations of the MNIST database of handwritten digits. 
\textbf{(a)}~This data is projected using TopoMap~\cite{doraiswamy2020topomap}. Note the presence of long branches with star-based cluster ensembles (colored) barely identifiable unless one zooms into the respective regions of interest.
\textbf{(b)}~The hierarchy defined by the process of topological simplification is visualized as a TreeMap. Each leaf of this tree corresponds to the smallest simplified component with a user\add{-}defined minimum number of points. There are eleven such components which are automatically selected as indicated by the colored borders. Each box in the TreeMap is colored-coded based on the persistence of the corresponding component.\add{The largest box is colored gray to indicate its infinite persistence.}
\textbf{(c)}~The TopoMap++ representation of the same data where the eleven components selected by the TreeMap are highlighted. As can be seen, TopoMap++ makes much more efficient use of the space compared to TopoMap, thus allowing users to easily analyze the relationships between the different clusters.
}
\label{fig:teaser}
}

\graphicspath{{figs/}{figures/}{pictures/}{images/}{./}} % where to search for the images

\usepackage{tabu}                      % only used for the table example
\usepackage{booktabs}                  % only used for the table example
\usepackage{lipsum}                    % used to generate placeholder text
\usepackage{mwe}                       % used to generate placeholder figures

\usepackage{mathptmx}                  % use matching math font

\usepackage{amssymb}
\usepackage{amsmath}
\newtheorem{lemma}{Lemma}
\newtheorem{definition}{Definition}
\usepackage{algorithm}
\usepackage{algpseudocode}
\algnewcommand{\IIf}[1]{\State\algorithmicif\ #1\ \algorithmicthen}
\algnewcommand{\EndIIf}{\unskip\ \algorithmicend\ \algorithmicif}
\usepackage{multirow}
\begin{document}

%%%%%%%%%%%%%%%%%%%%%%%%%%%%%%%%%%%%%%%%%%%%%%%%%%%%%%%%%%%%%%%%
%%%%%%%%%%%%%%%%%%%%%% START OF THE PAPER %%%%%%%%%%%%%%%%%%%%%%
%%%%%%%%%%%%%%%%%%%%%%%%%%%%%%%%%%%%%%%%%%%%%%%%%%%%%%%%%%%%%%%%


\maketitle

\section{Introduction}
\label{sec:intro}

Dimensionality reduction has long been a main visualization resource for analyzing and exploring high-dimensional datasets from various domains. 
Over the years, numerous dimensionality reduction methods have been proposed to translate high-dimensional data into a visual representation\hl{~\cite{Borg2006,cunningham2015linear,espadoto2019toward}}, most of which were designed to preserve geometric properties such as Euclidean distance between data points or distributions derived from it (e.g., see \hl{\cite{Maaten2007review,nonato2018multidimensional}}). An important challenge common to these methods is that the preservation of geometric properties can only be assured under specific conditions. Consequently, errors and distortions are highly probable in the resultant mapping. For example, structures present in the point cloud resulting from a projection, such as neighborhood relations, may not correspond to those in the original data, potentially misleading inexperienced practitioners and leading to incorrect conclusions. 

To enable more robust and reliable analysis, recent dimensionality reduction methodologies incorporate theoretical guarantees into the mapping process. Ensuring the preservation of topological properties is a main trend in this context (e.g., \hl{\cite{doppalapudi2022untangling,HWTopo10,Weber:2007}}).

In this work, we focus on one such approach, TopoMap~\cite{doraiswamy2020topomap}, which provides strong topological guarantees. Specifically, TopoMap ensures that the 0-cycles obtained by the Rips filtration of the projection are the same as in the original high-dimensional space.
However, the consequence of providing such a guarantee results in TopoMap having two weaknesses. First, it makes inefficient use of the visual space especially for large datasets. Second, the computational cost can be exorbitant when the data size and/or dimension is high.

This work presents TopoMap++, an adaptation of the TopoMap~\cite{doraiswamy2020topomap} algorithm that renders the TopoMap layout more effectively in terms of visual space usage.
The idea here is to understate the problematic points that are the cause of the inefficient visual space usage. 
We also provide a TreeMap~\cite{shneiderman1992tree} based exploratory mechanism that allows users to analyze high-dimensional data in a more robust and effective manner. The TreeMap is used to visualize the topological hierarchy of the high-dimensional data which allows for an intuitive exploration of the data set, thus making the analysis of the two-dimensional layout produced by TopoMap easier.
The layout improvement mechanism when combined with the TreeMap-based interactive exploration  facilitates the visualization of complex high-dimensional data structures by visually emphasizing the high-dimensional patterns found by the former. Note that, to the best of our knowledge, this the first projection approach that also allows for an interactive exploration.

Additionally, we present an approximation scheme that makes TopoMap more computationally efficient. This scheme drastically speeds up the most time-consuming step of TopoMap, the computation of the Euclidean minimum spanning tree.

In summary, the contributions of this work are:
\begin{itemize} \denselist
    \item \hl{TopoMap++, a layout improvement scheme to highlight important structures in the TopoMap layout where the structures are identified using the notion of topological simplification};
    \item A novel topology-guided TreeMap-based exploratory mechanism that facilitates the analysis of complex high-dimensional data;
    \item An approximation scheme that makes TopoMap more computationally efficient. We show that this approximation preserves the topology of the input while attaining at least two orders of magnitude speedup.
\end{itemize}

We also present case studies that demonstrate the effectiveness of our approach in analyzing several high-dimensional datasets.

\section{Related Work}

In order to better contextualize our contribution, we focus the related work discussion on techniques that explicitly rely on topology to perform dimensionality reduction. More comprehensive discussions can be found in
a number of surveys approaching different aspects of dimensionality reduction, including general overviews~\cite{engel2012survey,nonato2018multidimensional,ray2021various}, quantitative and qualitative comparisons~\cite{ayesha2020overview,espadoto2019toward,xia2021revisiting}, interaction tasks~\cite{sacha2016visual}, layout enrichment~\cite{sedlmair2013empirical,sohns2021attribute}, model specificities~\cite{ashraf2023survey,chao2019recent,cunningham2015linear,vu2022integrating}, and computational performance~\cite{velliangiri2019review}.

\emph{Isomap}~\cite{tenenbaum2000global} is a pioneering technique that employs topological mechanisms for dimensionality reduction. It resorts to a graph representation that captures the topological structure of the data and enables the estimation of geodesic distances. Variants of Isomap have emerged to speed up computation~\cite{silva2003global}, allow out-of-sample projections~\cite{bengio2004out}, and to handle spatio-temporal data~\cite{jenkins2004spatio}. Lee and Verleysen~\cite{lee2005nonlinear} improved the classical Isomap by building a graph representation that preserves non-contractable loops, enabling loop-preserving unfolding. Yan~et~al.~\cite{yan2018homology} target the preservation of cycles present in the original data by selecting landmarks based on a topological scheme that captures the structure of 1-dimensional homology groups, which are ideally preserved during the dimensionality reduction. However, 0-homology groups are not considered, which consequently remain unpreserved.
Gerber~et~al.~\cite{gerber2010, gerber2013} build upon Yan~et~al.~\cite{yan2018homology} to propose projection methods that are guided by a network derived from the maximum dimension cells of the Morse-Smale complex. However, rather than preserving topological structures, Gerber methods aim to encode information tailored for regression tasks.

The well-known UMAP~\cite{mcinnes2018umap} and t-SNE~\cite{van2008visualizing} techniques rely on KNN-graphs to capture the topological structure of high-dimensional data. While UMAP builds upon category theory and uses a force-directed scheme to project data to a visual space, t-SNE relies on the theoretical foundations of SNE~\cite{geoffrey:2002:nips}, minimizing the KL-divergence between distance distributions defined in the original and visual spaces. The main difference between t-SNE and SNE is the distance distribution defined in the visual space, which is assumed to be a t-student distribution in t-SNE. UMAP and t-SNE techniques have been widely used in a variety of applications, but they in no way guarantee the preservation of topological properties.
Doppalapudi~et~al.~\cite{doppalapudi2022untangling} proposed a topology-oriented force-directed layout that first generates an initial layout using the maximal spanning tree of a KNN graph of the high-dimensional data. Users then select $0$ and $1$-dimensional topological features in the corresponding persistence barcodes. Nodes belonging to selected $0$-dimensional cycles are attracted to emphasize the component while nodes in selected $1$-cycles are arranged in an elliptical shape using tailored forces.

Based on a terrain metaphor,  Weber~et~al.~\cite{Weber:2007} proposed a method to depict, in a two-dimensional layout, topological features of a 3D scalar field. Specifically, they carefully design a 2D terrain whose elevation contour tree is guaranteed by construction to match the original data's contour tree. However, Weber's method ignores metric information, thus it may project 3D topological features that are far apart in the original data space closer to each other in the 2D layout. In the same line, Harvey and Wang~\cite{HWTopo10} proposed a methodology to generate a terrain ensemble, each one with the same contour tree as the original data. Nonetheless, their method has the same shortcomings as Weber~et~al.'s~\cite{Weber:2007} approach.
Oesterling~et~al.~\cite{OesterlingHJS10} extended the terrain-based metaphor to high-dimensional data, relying on the Gabriel graph \cite{gabriel69} to build a simplicial representation of the data and on a kernel density estimation to derive a scalar field that faithfully captures the high-dimensional point cloud organization. The authors also proposed several improvements on the original approach~\cite{Oesterling0WS13,OHJSH11}, showing flexibility for application in different scenarios~\cite{OesterlingSTHKEW10}.  

The literature also brings topology-inspired regularization schemes to enforce soft topological constraints in the dimensionality reduction process~\cite{nelson2022topology,vandaele2021topologically,wagner2021improving}, but those methods do not provide any guarantee as to topological properties preservation. Topological tools have also been employed to evaluate and compare dimensionality reduction techniques~\cite{rieck2015persistent,rieck2017agreement}.

In contrast to the aforementioned techniques, the TopoMap~\cite{doraiswamy2020topomap} method offers theoretical guarantees as to the preservation of 0-dimensional homology groups, thereby ensuring that the connected components depicted in the projection layout match those in the original high-dimensional data. More specifically, TopoMap ensures, by construction, that the 0-dimensional persistence diagram of the Rips filtration of the projected data precisely mirrors that of the high-dimensional dataset. The topological guarantee provided by TopoMap renders it quite reliable as an analytical tool. In the present work, we tackle two main drawbacks of TopoMap, namely, computational cost and the inefficient use of visual space, thus substantially improving the analytical power of TopoMap. 
\section{Background: TopoMap}
\label{sec:bkgd}
In this section, we briefly introduce TopoMap and the related topological concepts. 
For an extended discussion over the topology topics, we refer the readers to Edelsbrunner and Harer's book~\cite{DBLP:books/daglib/0025666} and Chazal and Michel's introductory article~\cite{chazal2021introduction}. More details about TopoMap can be found in Doraiswamy~et~al.~\cite{doraiswamy2020topomap}.

Consider a set $P = \{p_1, p_2, \dots, p_n\}$ of $n$ high-dimensional points in $\mathbb{R}^d$. Let $\delta \geq 0$ be a threshold parameter. The \textbf{Vietoris-Rips complex}~\cite{DBLP:books/daglib/0025666} (also called Rips complex) $Rips_\delta(P)$ is the set of all simplices $K \subset P$, such that $d(p_i,p_j) \leq \delta$, $\forall p_i,p_j \in K$. 
Geometrically, given $\delta$, consider adding a $d$-dimensional ball of diameter $\delta$ around each point $p_i$. The Vietoris-Rips complex is equivalent to the set of all simplices formed by the points whose balls intersect. The threshold $\delta$ can be seen as the resolution we use to see the data set $P$~\cite{chazal2021introduction}.

The \textbf{Rips filtration} is defined as the nested sequence of subcomplexes $K_i$ formed by increasing $\delta$ from $0$ to $\infty$. It is an ordered set of subcomplexes $\mathbb{K} = \{K_0 = \emptyset, K_1, \dots, K_m\}$, where $K_i \subseteq K_{i+1}$ for all $i \in [0,m-1]$. Additionally, let $\delta_i$ be the smallest threshold such that $K_i \in Rips_\delta(P)$. Then, for all $i,j \in [0,m]$ with $i<j$, we have $\delta_i \leq \delta_j$.

As the threshold is varied, new simplices get added to the previous subcomplex. 
This addition can modify the topology of the subcomplex, where the topological features are $k$-dimensional cycles. Here, a $0$-cycle is a connected component, $1$-cycle is a loop, $2$-cycle is a void, and so forth. The topology is modified by creating or destroying one (or more) of such cycles. For a given $k$-cycle, we define $\delta_c$ and $\delta_d$ as thresholds at which the cycle is created and destroyed, respectively. The \textbf{topological persistence}~\cite{edelsbrunner2002topological} of the given $k$-cycle is the difference between $\delta_d$ and $\delta_c$. If a $k$-cycle is never destroyed, we define its persistence as infinite. 

The creation and destruction thresholds of the topological cycles are used to define the \textbf{persistence diagram}~\cite{cohen2005stability} of the Rips filtration. Specifically, the persistence diagram is a scatter plot where each point is a cycle created during the filtration, with the coordinates being the creation ($x$-axis) and destruction ($y$-axis) thresholds. The vertical distance between the point and the identity line ($x = y$) corresponds to the persistence of the cycle. We denote by $PD_P^k$ the persistence diagram containing only the $k$-cycles of the Rips filtration over the set $P$.

%With those definitions, we can formalize TopoMap's goal as: 
TopoMap's~\cite{doraiswamy2020topomap} projection approach tackles the following problem:
given a set $P=\{p_1,\dots,p_n\}$ in $\mathbb{R}^d, d > 2$, find a corresponding set of points $P' = \{p'_1, p'_2, \dots, p'_n\}$ in $\mathbb{R}^2$ such that $PD_P^0 = PD_{P'}^0$ with point correspondences between the $0$-cycles (that is, if $p_i$ belongs to a certain $0$-cycle in $PD_P^0$, then $p_i'$ must belong to the corresponding $0$-cycle in $PD_{P'}^0$). 
TopoMap accomplishes this by using the following result about the \textit{topology changing edges} (i.e., edges that merge two disconnected components into a single component) of the Rips filtration:

\begin{lemma}[Doraiswamy et al.~\cite{doraiswamy2020topomap}, Lemma 2]
    Let $\mathbb{K}_0 = \{e_1,e_2,\\\dots,e_{n-1}\}$ be the ordered set of topology changing edges of $P$. Then, $\mathbb{K}_0$ is exactly the set of edges of the Euclidean distance minimum spanning tree ($E_{mst}$) of the points $P$ in increasing order of length.
\label{lem:topomap}
\end{lemma}
where the Euclidean Minimum Spanning Tree is defined as follows:
\begin{definition}
Consider a set of points $P \in \mathbb{R}^d$, $|P|=n$. Let $K^n_E$ be the complete graph over P, where each edge has a weight equal the Euclidean distance between its end points. The \textbf{Euclidean Minimum Spanning Tree} $E_{mst}$ of $P$ is defined as the minimum spanning tree computed over the complete graph $K^n_E$.
\end{definition}

This lemma is then used to derive an iterative approach (Algorithm~\ref{alg:topomap}, black colored lines) to compute the projected points $P'$. 
The algorithm builds the projection one connected component at a time, where the connected components are processed in the order in which they are formed during the filtration defined by the $E_{mst}$ (see Figure~4 in \cite{doraiswamy2020topomap}). Note that this is equivalent to the Rips filtration, due to Lemma~\ref{lem:topomap}. 

\begin{algorithm}
\caption{TopoMap++}\label{alg:topomap}
\begin{algorithmic}[1]
\Require Set of points $P = \{p_1,\dots,p_n\}$;
\hlalgo{Set of Components $\C=\{C_1^*,C_2^*\,\ldots,C_k^*\}$}
\State Compute the Euclidean minimum spanning tree $E_{mst}$ of $P$
\State Let $E_{mst} = \{e_1,\dots,e_{n-1}\}$ be the edges ordered by length 

\color{blue}
\State Let $l_{max}$ be the maximum edge length in $E_{mst}$
\color{black}

\State $P' = \{p_1',\dots,p_n'\}$, where $p_i' = (0,0), \forall i$
\State Let $C_i = \{p_i'\}$ be the initial set of components
\For{each $i\in [1,n-1]$}
\State Let $(p_a, p_b)$ be the end points of edge $e_i$
\State Let $C_a, C_b$ be the components containing $p_a, p_b$, respectively
\State Let $l$ be the length of $e_i$
\State Place $C_a$ and $C_b$ in $\mathbb{R}^2$ s.t. $\min_{p_j'\in C_a, p_k'\in C_b} d(p_j', p_k') = l$
\State Let $C' = C_a \cup C_b$

\color{blue}
\If{$C' \in \C$}
    \State ScaleComponent($C'$, $l_{max}$)
\EndIf
\color{black}
\State Remove $C_a$ and $C_b$ from the set of components, and add $C'$
\EndFor
\end{algorithmic}
\end{algorithm}

\section{TopoMap++}
\label{sec:topomap-pp}

The projections generated by TopoMap are star-shaped ensembles with (often long) branches. 
Most of the points are usually concentrated quite densely in the center of such stars, with the branches taking up most of the visual space of the projections.
To analyze this projection, users must zoom into the centers of the different star shapes (e.g., see Figures~7, 8, and 9 of Doraiswamy~et~al.~\cite{doraiswamy2020topomap}).
This introduces two important issues. First, users can often miss features of interest if the star shapes are too small (relative to the area taken by the entire projection). 
Second, it is often hard to analyze multiple features at once, since when zooming into one, others often move out of frame.

This is primarily due to topological components with a single point (or components with very few points) that have high persistence during the Rips filtration. Such components spread the projection over the 2D space in order to maintain the topological consistency that is required by TopoMap.
A consequence of this is that larger components with similar or lower persistence take up a significantly smaller fraction of the area (used by the projection), thus making it difficult to easily identify such features.

Our goal in designing TopoMap++ is to allow such dense components to also be emphasized in the projection. The main idea is to first identify such components, which can then be reflected in the layout. At the same time, we also aim to provide users with flexibility in the exploration of the topological components.
In Section~\ref{sec:simplification}, we first describe how these components of interest can be identified using the notion of topological simplification.
We then present, in Section~\ref{sec:topomap-layout}, an alternate projection layout that adapts the original TopoMap layout to emphasize/highlight features of interest. This approach takes as input components of interest and generates a projection that focuses on the points present in these components. 
Finally, in Section~\ref{sec:treemap}, we propose using a TreeMap visual to allow users to interact and explore high dimensional data sets.

\subsection{Identifying Components of Interest}
\label{sec:simplification}

Consider the Euclidean minimum spanning tree $E_{mst}$ corresponding to a point set $P$. Let the edges of this tree be sorted in increasing order of edge weight. As mentioned in the previous section, these edges correspond precisely to the edges of the Rips filtration that merge two components (0-cycles). 
During this filtration, as each edge is added, new components are created by merging the two components that correspond to the end points of that edge. This creates a parent-child relationship between the components---the parent component is the union of the two child sub-components that are merged when an edge is processed. This can be represented as a tree, where each node corresponds to a component, and its children are the sub-components that are merged. Since only two components can merge for any edge, this tree becomes a binary tree.

\begin{figure}
\centering
\includegraphics[width=\linewidth]{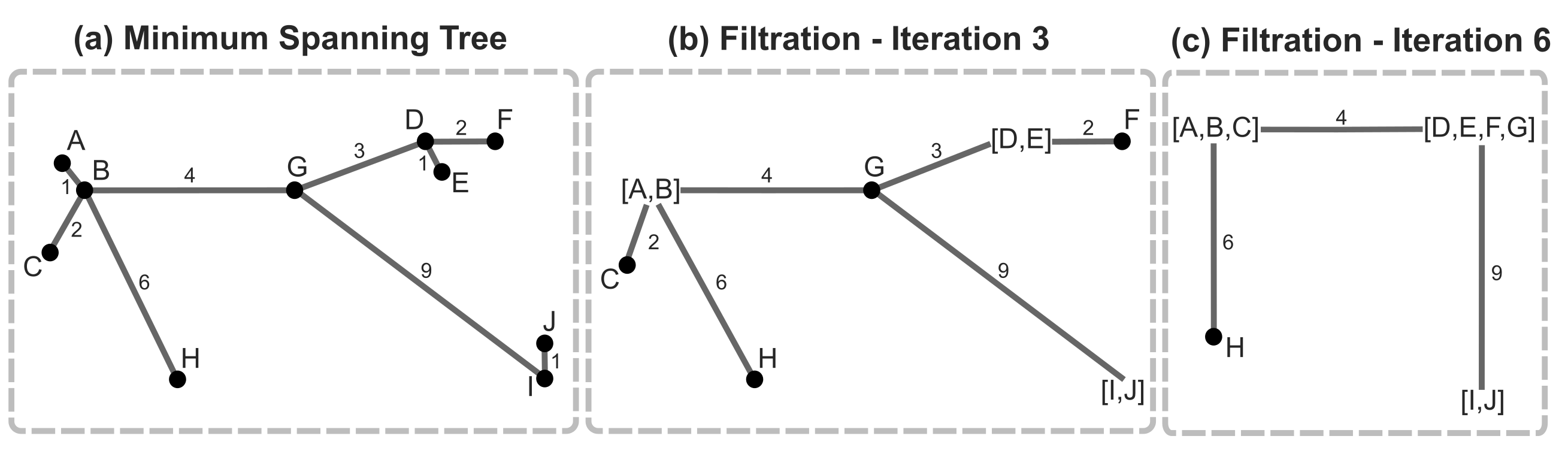}
\vspace{-0.1in}
\caption{
Computing the hierarchical tree based on the Rips filtration defined by the minimum spanning tree.
\textbf{(a)}~MST over an input with ten points (labeled from A to J). The edges weights (length) are specified for each edge.
\textbf{(b)}~The set of components after the filtration has processed 3 edges of the MST. The merged components now become a single node labeled using ``[]".
\textbf{(c)}~The set of components after the filtration has processed 5 edges of the MST..
}
\label{fig:tree-construction}
\vspace{-0.2in}
\end{figure}

\begin{figure}
\centering
\includegraphics[width=\linewidth]{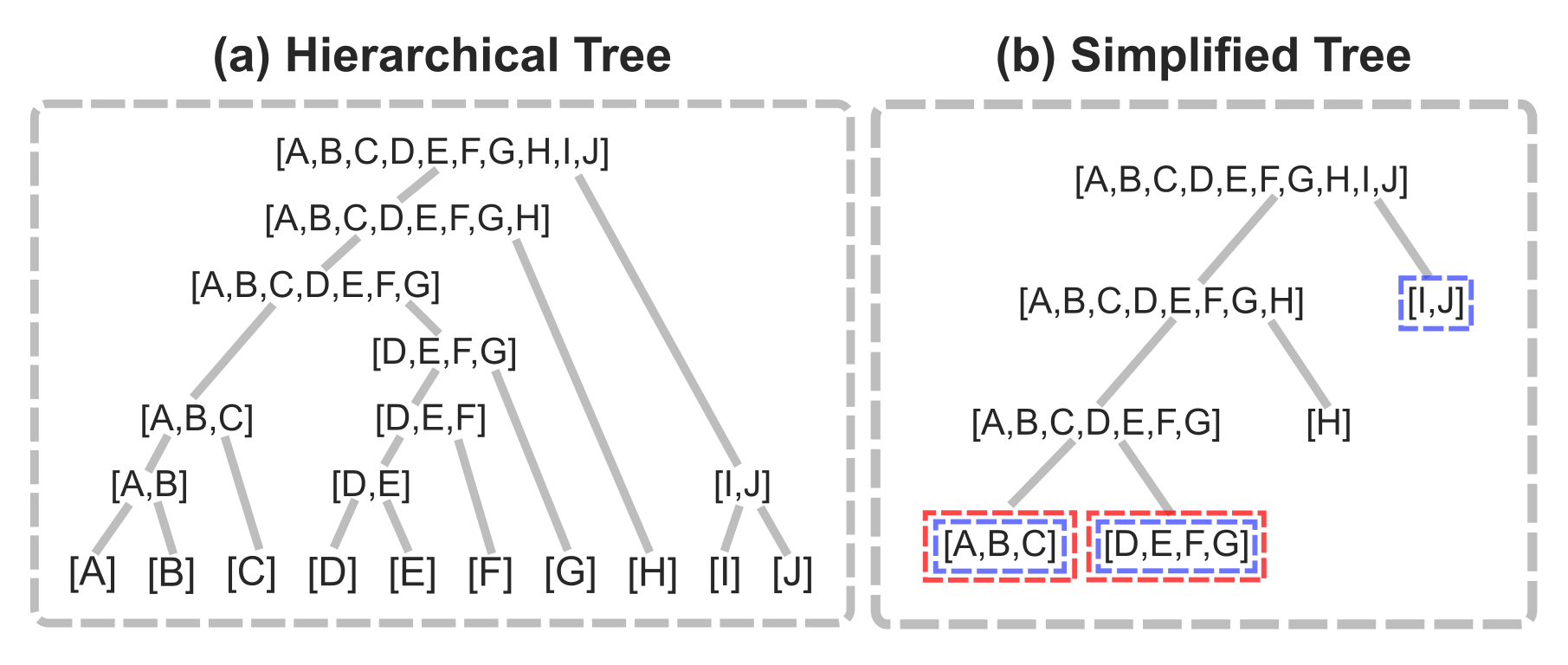}
\vspace{-0.2in}
\caption{
\textbf{(a)}~The hierarchy of the components formed during the filtration in Figure~\ref{fig:tree-construction} is represented as a hierarchical binary tree.
\textbf{(b)}~Simplified tree when $\eta=2$ (as well as when $\eta=3$. The components chosen when $\eta = 2$ are shown by a blue border, while those chosen when $\eta=3$ are shown by a red border.
}
\label{fig:tree-simplification}
\vspace{-0.1in}
\end{figure}

For example, consider an example $E_{mst}$ as shown in Figure~\ref{fig:tree-construction}(a). This was built on a point set $P$ with $|P| = 10$.
Figures~\ref{fig:tree-construction}(b) and (c) show the components that are created at different stages of the filtration.
The binary tree representing the hierarchy formed by the filtration is shown in Figure~\ref{fig:tree-simplification}(a). Each node in the binary tree corresponds to a component that is created during the Rips filtration. 

Our goal is to choose a disjoint set of components that are formed during the filtration to be highlighted during the projection. 
% These would essentially correspond to nodes in the binary hierarchical tree that do not share an ancestor-descendent relationship. 
%
Furthermore, we want to choose only components that are reasonably large, i.e., the size of the component (number of points) satisfies a minimum size criteria $\eta$. Thus, this set would not include the smaller high persistence components that are cause of the ``long branches".

We adapt the notion of topological simplification~\cite{Carr2004Simplification} to identify components that satisfy the above property.
Specifically, given a size threshold $\eta$, we simplify the hierarchical binary tree to identify only components that have size at least $\eta$.
This is accomplished by repeatedly merging leaf nodes in the binary tree that have size less than $\eta$ until no more leaf nodes can be merged. Note that a leaf node can be merged (simplified) if and only if the node it is merging with is also a leaf.
At the end of this procedure, each leaf node of the simplified hierarchical binary tree with size greater than or equal to $\eta$ corresponds to a component that contains a dense set of points as defined by the filtration. 
Figure~\ref{fig:tree-simplification}(b) shows the simplified tree, when $\eta = 2$ and $\eta = 3$, for the tree shown in Figure~\ref{fig:tree-simplification}(a). Note that, when $\eta=3$, the component [I,J] cannot be merged since its sibling node is not a leaf. However, when choosing the components of interest, this node is not considered.

\subsection{Space Efficient Layout}
\label{sec:topomap-layout}

The goal of our projection approach is enable the simultaneous highlighting of multiple features of the projection while still allowing users to visually infer the topological properties. 

Suppose that we have a list $\mathbf{C} = \{C_1^*, \dots, C_k^*\}$ of $k$ disjoint components of interest. That is, $C_i^* \bigcap C_j^* = \emptyset$, $\forall i \neq j$.
To highlight these components, our idea is to let these components take up more visual space when compared to other components. 
To accomplish this, we propose \textit{scaling} their projections to increase the area used by these components. 
Formally, for each component $C_i^*, i\in[1,\dots,k]$, our goal is to find a scalar $\alpha_i > 1$ such that for each $j \in C_i^*$, we replace $p'_j$ by $\alpha_i\cdot p'_j$ (wlog., assume that the center of this component is the origin). The TopoMap algorithm is modified to perform this scaling as soon as the component of interest is created (lines 11--13 in Algorithm~\ref{alg:topomap}).

\begin{figure}[t]
    \centering
    \includegraphics[width=0.7\linewidth]{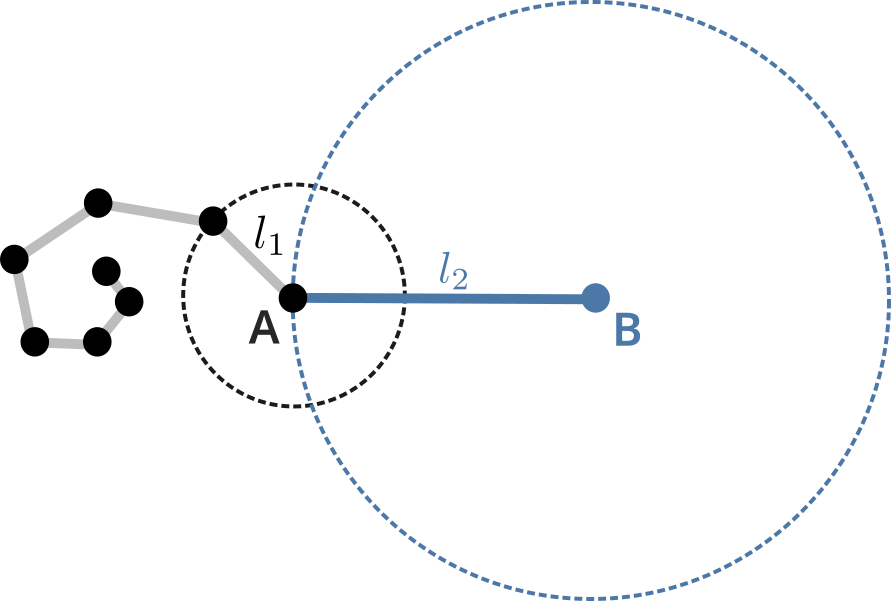}
    \vspace{-0.1in}
    \caption{Illustration of how the edge lengths impact the point density. The black points form a component and the segments correspond to the edges processed to form said component. Point A has two edges in the MST -- one connecting it to the black component (with distance $l_1$) and the other connecting it with point B ($l_2$). Since $l_1 < l_2$, A is the only point inside the ball centered at it with radius $l_1$. Similarly, B is the only point inside the ball around it with radius $l_2$. Since all edges between the points in the component are way smaller than $l_2$, their density in the visual space is higher than the density around point B.}
    \label{fig:densities}
    \vspace{-0.2in}
\end{figure}

To determine the scalar $\alpha_i$, we consider the edges in the MST that were processed to form the component. For a given data point $j \in [1,\dots,n]$, let $l_j$ be the length of the shortest edge in the MST containing $j$. In the projected space, there will be no other point inside the $2$-dimensional ball of radius $l_j$ centered in $p'_j$. Therefore, points with large shortest edges will be in sparse regions of the visual space, forcing the points with shorter edges to be cramped together (see Figure~\ref{fig:densities}). 
To highlight the components of interest, we scale the corresponding edges to increase their size, thus allowing the points to be placed farther apart. 
This is illustrated in Algorithm~\ref{alg:scale}.
Specifically, let $L_i, i \in [1,\dots,k]$ be the average edge length of component $C_i$. We would like $L_i$ to be at least as large as the biggest $l_j$ for any point $j$. 
In other words, we want $L_i \geq l_{max}$, where  $l_{max}$ is the longest edge of $E_{mst}$.
With that, we set the scaling factor $\alpha_i$ to be $\alpha_i = c\frac{l_{max}}{L_i}$, where $c \geq 1$ is a constant that can be used to control this scaling.
Note that we also allow users to optionally impose an upper bound on the scaling parameter $\alpha_i$ so that no single component takes up a disproportionate fraction of the visual space.

\begin{algorithm}
%\color{red}
\caption{ScaleComponent}\label{alg:scale}
\begin{algorithmic}[1]
\Require Component $C^*$; Max edge length $l_{max}$; Constants $c \geq 1$ and $\alpha_{max} \geq 1$ 

\State Let $L$ be the average edge length in $C^*$
\State Set $\alpha = \min \left\{c\frac{l_{max}}{L}, \alpha_{max}\right\}$

\For{$p \in C^*$}
\State Set $p = \alpha p$
\EndFor
\color{black}
\end{algorithmic}
\end{algorithm}

While this approach will no longer have the strong topological guarantees of the original TopoMap projection, it still retains the topological guarantees in the local neighborhood of the highlighted components. That is, the filtration corresponding to a highlighted component is the same (up to a constant scale factor) in both the input high-dimensional space as well as the project 2D space.

As an example, consider the MNIST dataset~\cite{mnist}  that is composed of images of handwritten digits. Figure~\ref{fig:teaser}(a) shows the original TopoMap projection of this data. As can be seen from the figure, the long branches in the projection result in an inefficient use of the visual space. 
The TreeMap++ layout that highlights the components that remain after simplifying the corresponding hierarchical tree using a value of $\eta$ set to be $1\%$ of the size of the dataset is shown in Figure~\ref{fig:teaser}(c).
Note the more efficient use of the visual space using this proposed layout.

\begin{figure}
\centering
 \includegraphics[width=\linewidth]{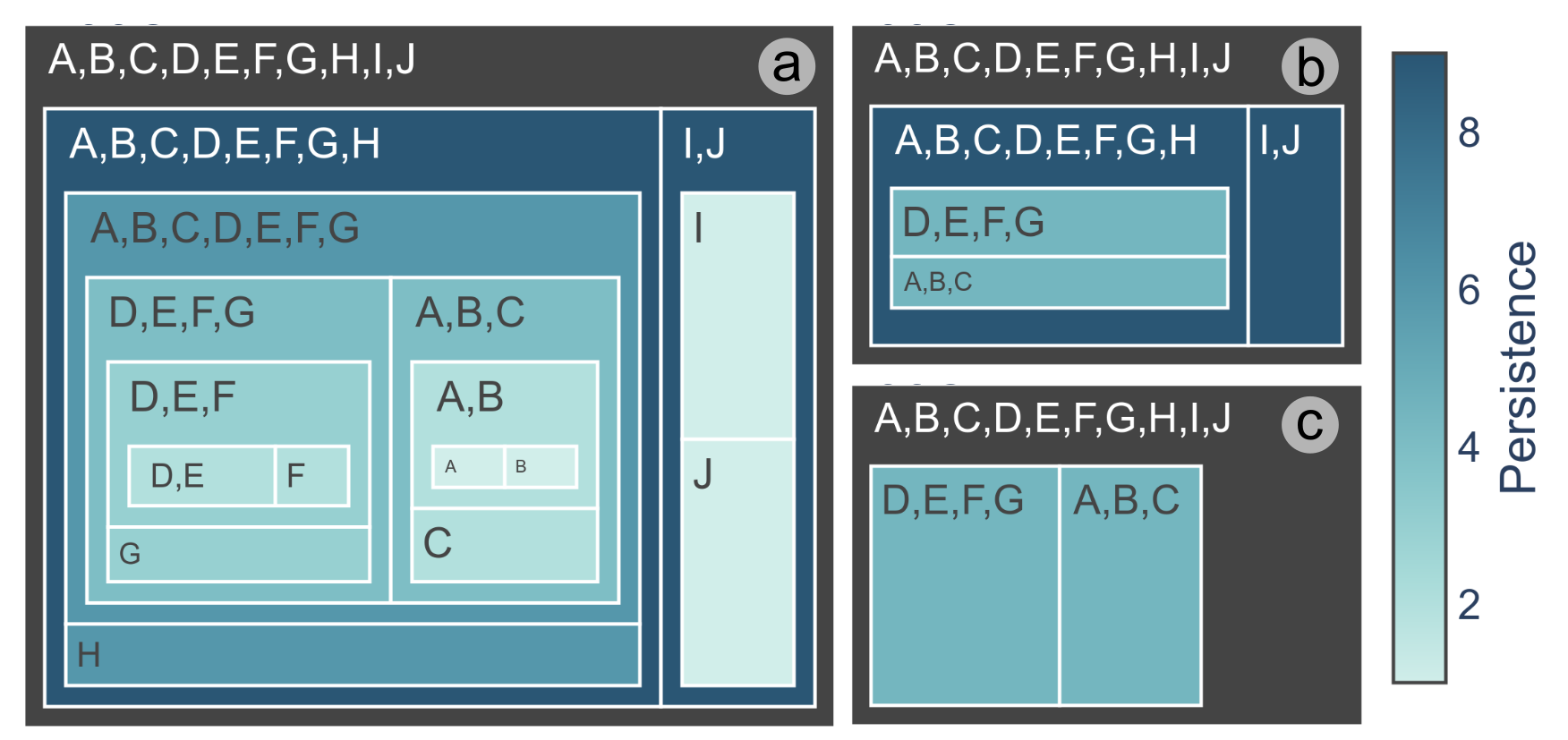}
 \vspace{-0.1in}
\caption{TreeMaps corresponding to 
\textbf{(a)}~the unsimplified hierarchical tree;
\textbf{(b)}~simplified tree with $\eta=2$; and 
\textbf{(c)}~simplified tree with $\eta=3$. The gray color represents the component with infinite persistence.}
\label{fig:treemaps}
\vspace{-0.2in}
\end{figure}

\subsection{TreeMap-based Exploration}
\label{sec:treemap}

The above approach chooses all components that satisfy the $\eta$ threshold. To provide more flexibility in data exploration to the user, we also aim to allow users to choose the components of interest. 
This allows the user to focus on fewer components, which in turn allows relatively more space to the components of interest.

To accomplish this, we propose to use the TreeMap~\cite{shneiderman1992tree} visual that can visualize the simplification hierarchy represented by the hierarchical binary tree.
Such a TreeMap provides an abstract hierarchical representation of the topology of the high-dimensional space. However, it can be visually cluttered if all components in the hierarchy are represented.
Thus, we choose to represent only the hierarchy present in the simplified binary tree to generate this visualization (using the $\eta$ parameter specified by the user).
We modify the original TreeMap layout to include boxes for a component \textit{if and only if} this component satisfies the threshold criteria. This enables users to focus only on components identified after simplification (see Figure~\ref{fig:treemaps}).
Furthermore, we color the boxes of the TreeMap to indicate properties of the components, such as persistence. 
By inspecting and interacting with the TreeMap, the user can define which component(s) to highlight in the projection.

\section{Approximate MST}

Recall that TopoMap's theoretical guarantees stem from the equivalence between the topology-changing edges in the Rips filtration and the Euclidean distance minimum spanning tree (EMST).
As mentioned in \cite{doraiswamy2020topomap}, the minimum spanning tree could also be computed using a different distance metric, such as cosine distance, and used for the projection. 
For the rest of this section, we assume that the distance metric is the Euclidean distance. 

Computing the EMST, however, can be prohibitively expensive for large and/or high-dimensional datasets. 
To address this problem, we propose to use an approximation of the EMST instead of the actual EMST. To do this, our idea is to first reduce the complete graph defined by the input points to a substantially smaller subgraph $G'$ and then compute the MST of $G'$. 
The key-insight to compute this smaller subgraph $G'$ is inspired by the state-of-the-art approximate nearest neighbor (ANN) algorithms. In this section, we first describe the Vamana graph which provides us with the necessary subgraph $G'$ to compute the approximate EMST. 
We then evaluate this approximation, demonstrating its effectiveness with respect to the quality of approximation and its efficiency compared to computing the exact EMST.

\subsection{Vamana graph}

A \textit{Relative Neighborhood Graph} (RNG) is an undirected graph constructed on a set of points $P$ such that there is an edge between two points $u,v$ if and only if there is no point $p \in P$ that is closer to both $u$ and $v$ than they are to each other. 
More formally, for a set of points $P$ in a metric space with distance $d$, the RNG of $P$ is a graph with vertex set $P$ and set of edges equal to those pairs $(u,v)$ such that $d(u,v) \leq max_{p \in P \backslash \{u,v\}}(d(u,p), d(v,p))$. 
An RNG allows for an efficient identification of the nearest neighbors corresponding to a query point. 
However, since computing the RNG is expensive, especially for higher dimensional data sets, several ANN algorithms use an approximate variant of the RNG.

By definition, an EMST is a subgraph of the RNG. Given this, our idea is to use the 
\textit{Vamana graph}~\cite{NEURIPS2019_Vamana}, which is a sparse approximation of the RNG, to compute an approximate EMST.
For completeness, we briefly describe the Vamana graph construction algorithm next. We refer the reader to the DiskANN paper~\cite{NEURIPS2019_Vamana} for more details.

The Vamana indexing algorithm constructs the graph iteratively as follows.
It takes 3 parameters as input: a distance threshold factor $\alpha$ that determines the diameter of the approximate RNG; an upper bound on the out degree of each node, $R$; and the allowed search list size $L$ used for doing a greedy ANN search during the graph build process.
It first computes a random $R$-regular directed graph $G'$ over the input set of points $P$. 
Then, during each iteration, it searches for the approximate nearest neighbors of a random point in $p_i \in P$ that is not yet processed and updates the out-neighbors of this point $p_i$ based on the search results. Additional pruning is done during each step to ensure that the out-degree of the graph $G'$ is within the upper bound $R$.

To compute the EMST, we first compute the Vamana graph $G'$ and then compute the EMST using this graph.

\begin{table}[t]
\footnotesize
\caption{Data sets used in our experiments.}
\label{tab:datasets}
\vspace{-0.2in}
\begin{center}
\begin{tabular}{|l|c|c|c|}
\hline
\textbf{Data set} & \textbf{\# Points} & \textbf{Dimension} & \textbf{\# Classes} \\ \hline 
Iris \cite{Dua:2019UCI_datasets} & 150 & 4 & 3 \\ \hline
Seeds \cite{Dua:2019UCI_datasets} & 210 & 7 & 3 \\ \hline
Mfeat \cite{Dua:2019UCI_datasets} & 2000 & 64 & 10 \\ \hline
MNIST~\cite{mnist} & 60000 & 784 & 10 \\ \hline 
BIGANN~\cite{BIGANN_DATASET} & 100000 & 128 & not labeled \\ \hline \hline
LLM & 6669 & 4096 & 2 \\ \hline
Urban & 17520 & 6 &  not labeled \\ \hline
StreetAware& 363134 & 768 & 3 \\ \hline
\end{tabular}
\end{center}
\vspace{-0.2in}
\end{table}

\subsection{Evaluation}

To compute the approximate EMST, denoted as \textit{AMST}, we first compute the Vamana Graph $G'$ with  $\alpha = 1.3$, $L = 100$ and $R = 100$ and then extract the MST of $G'$. We found that this set of parameters provided a good tradeoff between accuracy and running time.
% , as implemented in Python's \textsc{scipy} package \cite{2020SciPy-NMeth}. 
%
The exact EMST used for comparison was computed using the Dual-Tree algorithm~\cite{10.1145/1835804.1835882/FastEMST} that was implemented as part of the  \textsc{mlpack} library~\cite{mlpack2023}. Note that this was the same implementation that was also used in \cite{doraiswamy2020topomap}.

Table~\ref{tab:datasets} presents the number of data points, dimensions, and classes of the datasets used in our evaluation. \add{The first 5 datasets are well known open datasets and are used purely for the quantitative evaluation, while the last three datasets are used for both the  quantitative evaluation as well as the case studies discussed in the next Section}. All experiments were run on a machine with Intel (R) Core(TM) i9-12900KF running at 3.19GHz and 32 GB of memory. 

\subsubsection{Efficiency}

Table~\ref{tab:Time_Comparison} shows the performance improvement of our proposed approach for computing the AMST. We note that for small low-dimensional datasets, building the Vamana graph incurs a small overhead. However, given that the total running time itself is very small, this is insignificant. However, as the data sizes/dimensions increase, we note that our approach provides a significant speedup over computing the exact MST, attaining over \textbf{two orders of magnitude} speedup.

\begin{table}[t]
\footnotesize
\centering
\caption{Comparing time to compute the EMST  and the AMST. Note that as the data size/dimension increases, we are able to achieve a significant speedup in the running times.
}
\label{tab:Time_Comparison}
\vspace{-0.1in}
\begin{tabular}{|l|c|c|c|} \hline
\multirow{ 2}{*}{\textbf{Dataset}} & \multicolumn{2}{c|}{\textbf{Running Time (sec)}} & \multirow{2}{*}{\textbf{Speedup}} \\ 
\cline{2-3}
        & \textbf{EMST} & \textbf{AMST} &       \\ \hline
Iris    & 0.003         & 0.02          & 0.15  \\ \hline
Seeds   & 0.002         & 0.012         & 0.17 \\ \hline
MFeat   & 0.332         & 0.302         & 1.1   \\ \hline
MNIST   & 9010          & 28            &321.8 \\ \hline
BIGANN  & 2317          & 20            &115.8  \\ \hline \hline
LLM & 542      & 1.3           &416.9  \\ \hline
Urban  & 0.18          & 1.03            &0.17  \\ \hline
StreetAware& 92631      & 156           &593.8  \\ \hline
\end{tabular}
\vspace{-0.1in}
\end{table}

\subsubsection{Approximation Quality}

We use the following two metrics to evaluate the approximation quality of our AMST approach:
\begin{enumerate} \denselist
\item \textbf{Bottleneck Distance:}
This measure is used to assess the topological similarity between the filtration defined by the AMST and the EMST (i.e., the original Rips filtration). 
It is computed as the bottleneck distance between the persistence diagrams defined by the two filtrations. Note that during these computations, the persistence diagrams are normalized to ignore any scaling effects.

\item \textbf{Relative Weight Error (RWE):}
This measure is used to assess the quality of the approximation attained by the AMST. It is defined as the difference between the total weights of the AMST and EMST normalized by the weight of the EMST: 
$\text{RWE} (\text{AMST},\text{EMST}) = \frac{W(\text{AMST}) - W(\text{EMST})}{W(\text{EMST})}$.
Here, $W(.)$ is the total weight of a given weighted tree.
\end{enumerate}

\hide{
\begin{table}[t]
\centering
\caption{Approximation quality of the AMST.}
\label{tab:acc_comparison}
\begin{tabular}{|l|c|c|c|}
\hline
\multirow{ 2}{*}{\textbf{Dataset}}     & \multicolumn{2}{c|}{\textbf{\add{Distance}}} & \multirow{2}{*}{\textbf{RWE}} \\ 
\cline{2-3}
        & \textbf{Bottleneck} & \add{\textbf{Wasserstein}} & \\ \hline
Iris       & 0  & \add{0} &  0 \\ \hline
Seeds       & 0 & \add{0} &  0\\ \hline
Mfeat       & 0 & \add{0} & 0 \\ \hline
MNIST       & 2.4$\times 10^{-2}$ & \add{1332.3}& 1.86 $\times 10^{-4}$ \\ \hline
BIGANN & 2.6 $\times 10^{-2}$ & \add{1762.8}& 6.39 $\times 10^{-4}$ \\ \hline \hline
LLM &  $ 7.5 \times 10^{-2}$ & \add{23}& 2.01 $\times 10^{-3}$ \\ \hline
Urban &  $ 1.3 \times 10^{-2}$ &\add{0.1}& 4.55 $\times 10^{-3}$ \\ \hline
StreetAware & 6.0 $\times 10^{-2}$&\add{7442.9} & 7.74 $\times 10^{-4}$ \\ \hline
\end{tabular}
\end{table}
}

\begin{table}[t]
\centering
\footnotesize
\caption{Approximation quality of the AMST.}
\label{tab:acc_comparison}
\vspace{-0.1in}
\begin{tabular}{|l|c|c|} 
\hline
\textbf{Dataset}     &  \textbf{Bottleneck Distance} & \textbf{RWE} \\  \hline
Iris       & 0  &  0 \\ \hline
Seeds       & 0 &  0\\ \hline
Mfeat       & 0 & 0 \\ \hline
MNIST       & 2.4$\times 10^{-2}$ & 1.86 $\times 10^{-4}$ \\ \hline
BIGANN & 2.6 $\times 10^{-2}$ & 6.39 $\times 10^{-4}$ \\ \hline \hline
LLM &  $ 7.5 \times 10^{-2}$ &  2.01 $\times 10^{-3}$ \\ \hline
Urban &  $ 1.3 \times 10^{-2}$ & 4.55 $\times 10^{-3}$ \\ \hline
StreetAware & 6.0 $\times 10^{-2}$& 7.74 $\times 10^{-4}$ \\ \hline
\end{tabular}
\vspace{-0.1in}
\end{table}

Table \ref{tab:acc_comparison} compares the above two metrics for the different datasets shown in Table~\ref{tab:datasets}. 
Note that for all datasets, the relative weight error between the AMST and EMST is very small (in the order of $10^{-1}\%$ or less). We notice that, for smaller datasets, the AMST is exactly the same as the EMST (\textsc{RWE} = 0). 
Even in cases where the AMST does not match the EMST, we notice that the bottleneck distance between the persistence diagrams is still small, thus ensuring that the topology of the projection is still mostly preserved with significantly lower computational effort. 

\section{Case Studies}

In this section, we first discuss the layout and interpretation of TopoMap++. We then describe three use case scenarios that uses TopoMap++ to analyze datasets. Note that, unless otherwise mentioned, we use $\eta = 1\%$ of the dataset size as the simplification threshold.

\subsection{TopoMap++: Layout and Interpretation}
\label{sec:interpretation}

TopoMap's layout consists of star-shaped ensembles with branches connecting and emanating from them. The center of the star shapes are denser than the branches and contain points that are closer together in the original space. 
These correspond to dense topological components that could be of interest in the analysis.
For example, the centers in Figure~\ref{fig:teaser}(a) make up points of the same digit. 
 
Since the projected points satisfy the topological filtration guarantee, the branches connecting any two such components indicate the order in which these components merge during the filtration. This could be used to understand the ``connection relationship" -- the points that are responsible for connecting these components. 
Returning to the MNIST example, we see that such a branch connecting two components typically contains points that gradually change patterns between the components -- in this case we see a transition between the digits that are connected (e.g., see Figure 7 of Doraiswamy~et~al.~\cite{doraiswamy2020topomap}). 

A common trait seen in various datasets is that they often also contain several small components (sometimes with just 1 point) that have high persistence.
When there are a lot of high-persistent small-sized components, the layout tends to make inefficient use of the visual space. This makes it difficult to 
(1)~identify components of interest; and 
(2)~understand the relationship between these components.
Both these shortcomings are addressed using TopoMap++, as we show next using the case studies.
Using TopoMap++, selected components are enlarged and may be emphasized with colored regions in the projection plot.
Since we make use of the notion of topological simplification to identify components of interest, TopoMap++ allows for a better analysis of the topological components themselves.

In the MNIST example, we can barely notice the presence of most of the star centers using the TopoMap layout~(Figure~\ref{fig:teaser}(a)). In contrast, the TopoMap++ projection (Figure~\ref{fig:teaser}(c)) more clearly displays these centers, 
thus allowing the user to see all of the points that form each of these components.
At the same time, since such components are already identified and emphasized using TopoMap++, it now becomes possible to also study the relationship between them much more easily.
However, care should be taken while analyzing TopoMap++ layout to remember that
the filtration (distances up to a constant scaling) is consistent within each emphasized component but not across them.

\begin{figure*}[!t]
\centering
\includegraphics[width=\linewidth]{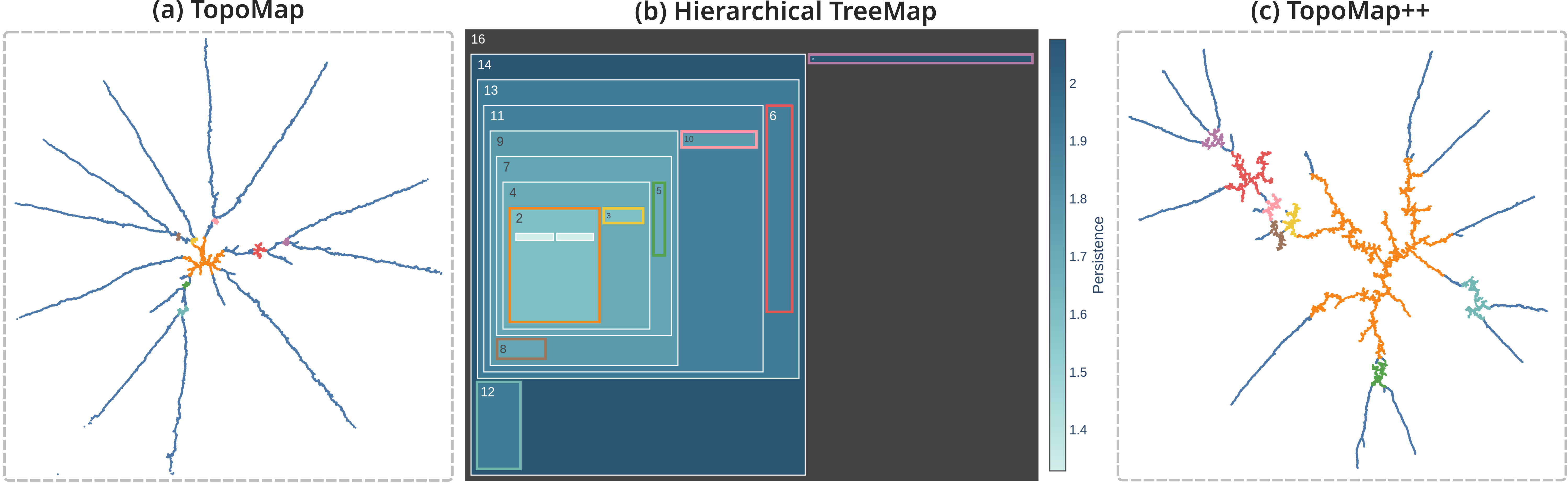}
\caption{\textbf{(a)} TopoMap projection, \textbf{(b)} Hierarchical TreeMap, and \textbf{(c)} TopoMap++ projection generated using the urban data set from Case Study 1. The selected components in \textbf{(b)} are used to emphasize the clusters in the TopoMap++ projection \textbf{(c)}. The same points are also colored with the corresponding colors in the original TopoMap projection \textbf{(a)}. Note that these components end up being small due to the inefficient use of the visual space by the original algorithm. Using our proposed approach, it becomes easy to identify and analyze such features in the data.}
\label{fig:CaseStudy-Urban}
    
\includegraphics[width=\linewidth]{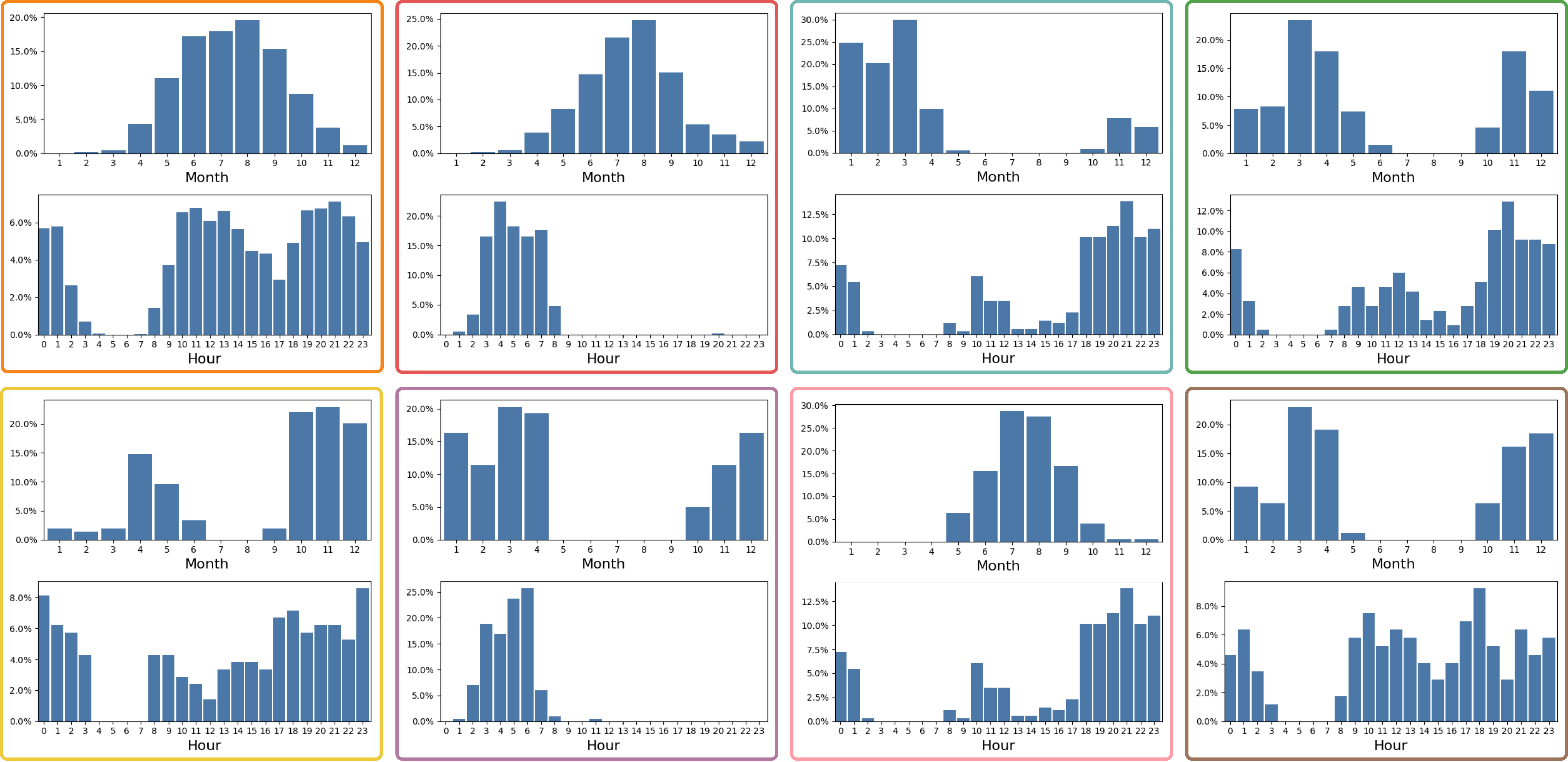}
\vspace{-0.1in}
\caption{Histograms of the temporal components (month and hour) for each cluster highlighted in Figure~\ref{fig:CaseStudy-Urban}. The colored boxes correspond to these clusters. The components are ordered (from left to right, top to bottom) according to their volume (i.e., number of points). }
\label{fig:CaseStudy-Urban-Histograms}
\vspace{-0.1in}
\end{figure*}

\subsection{Case Study 1: Unlabeled Urban Data}

In the first case study, we directly compare TopoMap++ with TopoMap for the same unlabeled urban data set used in Doraiswamy~et~al.~\cite{doraiswamy2020topomap}(Section~4.2), with the goal to assess:
(1)~the ease of exploring the projection using our proposed approach; and
(2)~the quality of the features identified/explored when compared to a manual exploration.

The data set contains six features for a 100-meter radius region in Times Square (precipitation, temperature, wind speed, count of taxi pickups, average fare, and average distance) for hourly intervals during 2014 and 2015 (for a total of 17,520 six-dimensional data points). In Doraiswamy~et~al.~\cite{doraiswamy2020topomap}, this data set is manually explored through the TopoMap projections. Here, we show that TopoMap++
significantly improves the unlabeled data exploration compared to TopoMap.

As discussed earlier, generating the TopoMap++ projection first computes and simplifies the hierarchical binary tree. This layout is shown in 
Figure~\ref{fig:CaseStudy-Urban}(c), where each component has a different color. 
The corresponding TreeMap is shown in Figure~\ref{fig:CaseStudy-Urban}(b).
Figure~\ref{fig:CaseStudy-Urban}(a) shows the TopoMap projection for the same dataset in which we can barely see some of the components (for example, those in pink and gray) --- these components could easily be missed during the visual exploration. 

\begin{figure*}[t]
\centering
\includegraphics[width=\linewidth]{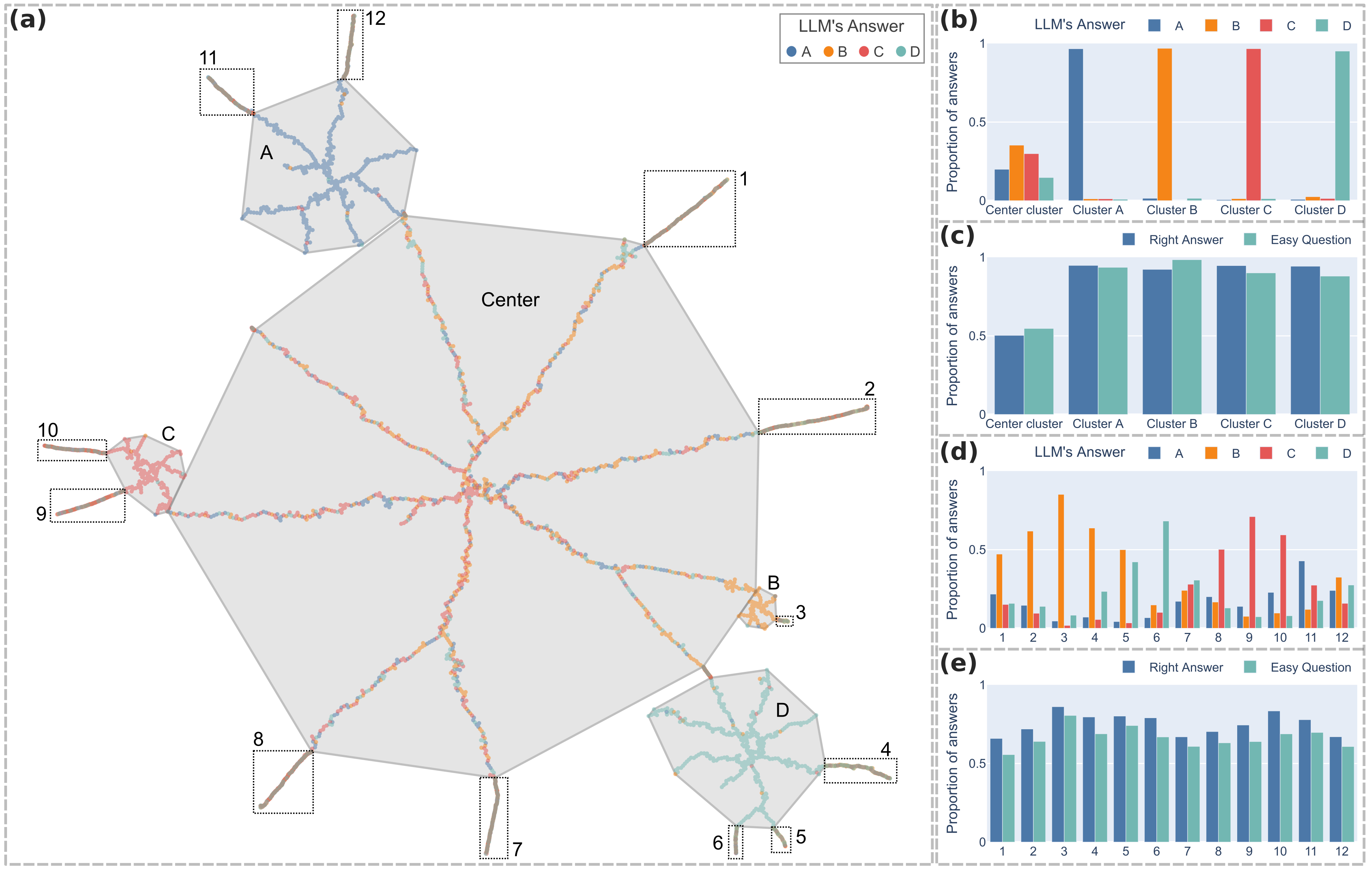}
\vspace{-0.1in}
\caption{(a) TopoMap++ projection of the LLM embeddings dataset explored in Case Study~2. The grey regions indicate the convex hull of the five highlighted components. 
There are also twelve branches besides the highlighted components, which are indicated with bounding boxes and are numbered clockwise.
The proportion of each option answered by the LLM is shown in (b) and (d) for the highlighted components and the numbered branches, respectively.}
\label{fig:CaseStudy-LLM}
\vspace{-0.2in}
\end{figure*}

\addtocounter{figure}{1}
\begin{figure*}[t]
\centering
\includegraphics[width = \linewidth]{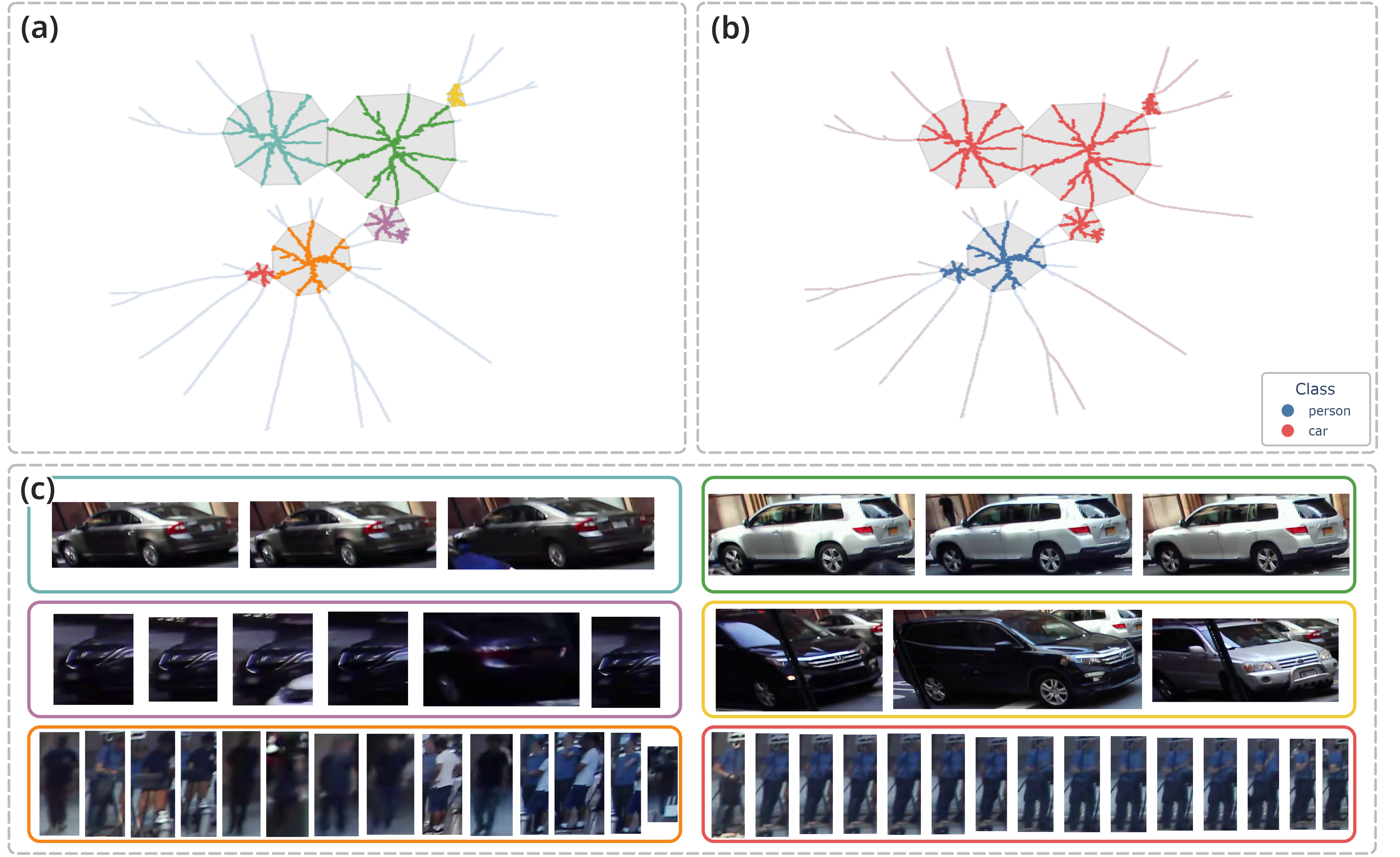}
\vspace{-0.1in}
\caption{TopoMap++ projections of the Street Aware embeddings data set. The grey regions indicate the convex hull of the highlighted components. 
\textbf{(a)}~Projection the StreetAware data using TopoMap++ highlighting six prominent components.
\textbf{(b)}~The points in the projection are colored based on the YOLOv8 classification label. Note that even within the same class, there are sub-clusters. 
\textbf{(c)}~Sample images corresponding to the different highlighted clusters are shown. We can see that the sub-clusters within a given class correspond to different variations of the data.}
\label{fig:CaseStudy-StreetAware}
\vspace{-0.2in}
\end{figure*}

Next, we analyze how well the automatically selected components compare to the features found manually by Doraiswamy~et~al.~\cite{doraiswamy2020topomap}.
Figure~\ref{fig:CaseStudy-Urban-Histograms} shows the histograms of the month and hour of the day for each of the eight components. Most of the patterns found in \cite{doraiswamy2020topomap} match the components found automatically here: 
the biggest component (in orange) corresponds to intervals during the spring and summer months during the day (between 8 am and 2 am). This pattern corresponds to the points shown in \cite{doraiswamy2020topomap}, Fig.~8(b);
the component in red corresponds to night intervals from spring and summer (matching cluster \cite{doraiswamy2020topomap}, Fig.~8(f)); the teal component contains day intervals during the winter (matching \cite{doraiswamy2020topomap}, Fig.~8(d)); yellow has day intervals during the spring and fall (matching \cite{doraiswamy2020topomap}, Fig.~8(e)). 

Beyond matching the clusters from \cite{doraiswamy2020topomap}, the automatic selection proposed in this paper also found additional patterns. 
The green component finds mostly the night hours during Winter (with a few points bleeding from Fall and into Spring). 
The purple one contains the periods during the early hours of the day after midnight for the fall and winter months, complementing the red component in terms of season. The pink component has summer periods during the night (mostly before midnight) when there are more taxi pickups than usual during the night. 
This ability to quickly and easily find features of interest can greatly help in the analysis of high dimensional datasets.

In addition to easily finding such interesting components, we can now further examine their structure, such as neighborhoods, to better understand the relationship between them. As mentioned above, components closer merge earlier in the filtration. 
For example, we notice that the red component has two components in its neighborhood--pink and purple. The pink component corresponds to the same monthly intervals as the red one, while differing in the hours of the day that these points represent.
On the other hand, the purple component represents the same hours of the day as the pink component while differing in the months (seasons) that it represents.

Note that the only component from Doraiswamy~et~al.~\cite{doraiswamy2020topomap} that was not automatically identified here is the set of periods with rainfall (\cite{doraiswamy2020topomap}, Fig.~8(g)). 
This component has a small size but is high persistent, which is basically the type of component that gets simplified by our approach. While such components can be obtained by manually exploring the projection, in future, it would be interesting to identify alternate approaches to automatically highlight such components as well.

\subsection{Case Study 2: Analyzing LLM Embeddings}

The analysis of high dimensional embeddings (or hidden states) of large machine learning models, in particular large language models~(LLM), is a current hot topic in high dimensional analysis. By analyzing these embeddings, we can further understand how those models interpret text, answer questions, and perform tasks. In this case study, we employ TopoMap++ to explore patterns in the embedding space of a large language model when answering multiple-choice questions. 
We use the AI2 Reasoning Challenge (ARC) data set~\cite{Clark2018ThinkYH}, which contains grade-school level questions split into Easy and Challenge sets. To answer those questions, we used the Llama~2 model with $7$B parameters~\cite{touvron2023llama}. 
Ten sample questions were provided, with answers, to the model as examples (i.e., a fewshot learning). 
Then, we asked the model the questions from the ARC question set and extracted the embeddings that resulted in the model's answer. In particular, we looked at the embeddings after the last attention layer of the model. For each of the 6,669 questions\footnote{We selected the ARC questions whose options were letters (A, B, C, or D) from the Train and Test sets for the Easy and Challenge levels. The examples provided were randomly selected from the Dev sets.}, we have one embedding of dimension 4,096. For reference, $71.6\%$ of the questions were answered correctly, with an accuracy of $77.6\%$ for the Easy set and $59.7\%$ for the Challenge set. Also, Table~\ref{tab:llm-answers} shows the number of correct answers of each possible choice and the number of answers the model gave.

\begin{table}[t]
\footnotesize
\centering
\caption{Number of correct and model answers from each option.}
\label{tab:llm-answers}
\vspace{-0.1in}
    \begin{tabular}{|c|c|c|}
        \hline
        \textbf{Choice} & \textbf{Correct Answer} & \textbf{Model Answer}\\
        \hline
        A & 1,589 & 1,411\\\hline
        B & 1,721 & 2,066\\\hline
        C & 1,745 & 1,785\\\hline
        D & 1,614 & 1,407\\\hline
    \end{tabular}
\vspace{-0.2in}
\end{table}

We computed the TopoMap++ projection for this data set setting $\eta$ as $50$ ($~1\%$ of the data size)--this resulted in five large disjoint components as shown in Figure~\ref{fig:CaseStudy-LLM}(a).
Since the questions had four possible answer options, we first explored these five components to assess how they associate with the four answer options. 
Here, the highlighted components are enclosed within grey convex hulls, and
the points are colored based on the LLM's answer. 
In Figure~\ref{fig:CaseStudy-LLM}(b), we see that each of the four outer components corresponds to one of the four choices made by the LLM. 
The fifth component is a central component whose points correspond to a mix of all four answer options but with more B and C answers than the others.

We then analyzed whether the LLM provided the correct answer or not for the questions in each of the five components, as shown by the blue bars in Figure~\ref{fig:CaseStudy-LLM}(c). It is interesting to note that all the answers made by the LLM for questions in the outer four components were correct, while the central region consists of questions where the LLM mostly answered incorrectly. The latter also holds for the points in the long branches connected to this central component. Figure~\ref{fig:CaseStudy-LLM}(c) also shows, as green bars, the proportion of questions from the Easy set for each component.
Note that this is very close to the proportion of correct answers, implying that 
each outer component corresponds to only easy questions.

Furthermore, the presence of this central component indicates a contiguous region in the higher dimensional space where the LLM frequently makes mistakes. 
Based on the projection, it is also clear that this region separates other regions that mostly answer correctly. This knowledge can potentially be used for improving both the answer quality (e.g., by augmenting the inference with additional information when the embedding falls within such region), as well as fine tuning the LLM by focusing on such regions. 

Beyond the components themselves, the projection in Figure~\ref{fig:CaseStudy-LLM}(a) also displays twelve branches that are not contained by any component. Four of those branches (numbered $1, 2, 7,$ and $8$) are directly connected to the central component, while the other eight are connected to one of the outer components. In Figure~\ref{fig:CaseStudy-LLM}(e), we see the proportion of correct answers (in blue, ranging from $66\%$ to $86\%$) and Easy questions (in green, ranging from $55\%$ to $80\%$) for each branch. Compared to the central component, all branches have higher accuracy and more Easy questions. However, those values are smaller than the ones corresponding to the outer components. The branches directly connected to the central component show a slightly lower accuracy and also contain a smaller proportion of Easy questions than the other branches.

Lastly, Figure~\ref{fig:CaseStudy-LLM}(d) shows the proportion of answer options in each branch. We notice that the branches connected to an outer component tend to have more answers of the corresponding option (branch 3 has mostly B answers, branches 6 and 7 mostly D, branches 9 and 10 mostly C, and branch 11 mostly A). However, the branches close to the B component (from 1 to 4) have mostly B answers, even though branch 4 is connected to component D. A similar pattern happens from branches 8 to 10, which have mostly C answers. The only branches with mostly A or D answers are branches 11 and 6, respectively. This is consistent with our previous observation that the B and C answers are scattered across the embedding space instead of being closer together as we see with answers A and D. This might indicate that the LLM might be biased towards choices B and C, and they therefore become the ``default" choices when the model does not know the answer.

\subsection{Case Study 3: Vision Transformer Embeddings}

Examining the embeddings of Vision Transformer~(ViT) provides useful information on how neural networks handle visual data, assisting in the identification of objects and comprehension of scenes. Understanding these embeddings elevates the transparency and durability of the model, which is critical for dependable AI systems. In this case study, we use TopoMap++ to analyze the embeddings of ViT~\cite{Dosovitskiy2020AnII}. 
For this study, we use the StreetAware dataset~\cite{piadyk_streetaware_2023}, which is a dataset with fully anonymized high-resolution videos from urban environments. From this dataset, we selected a 46-minute video stream that was recorded using a fixed camera in an activity-rich traffic intersection and then utilized the YOLOv8~\cite{yolov8_ultralytics} detection model to detect bounding boxes for each object identified as `car' or `person' in every video frame. 
We then crop the images using these bounding boxes and compute the ViT model embedding for each of the cropped images. So, for each ViT embedding, there's a YOLOv8 score and label. 

The TopoMap++ projection generated by using a value of $\eta = 3600$ (approximately $1\%$ of the number of data points) resulted in highlighting 6 components as shown in Figure \ref{fig:CaseStudy-StreetAware}(a).
Figure~\ref{fig:CaseStudy-StreetAware}(b) shows the same visualization but with the points colored based on its labels. Note that each component corresponds to a single label. 
Interestingly, each label is split into multiple sub-components. To analyze this, we look at the images corresponding to these points.
Figure~\ref{fig:CaseStudy-StreetAware}(c) contains sample images from each sub-component 
where we can see that these components form a pattern within each category. 
For example, the cyan and green components consist of embeddings related to two static cars, respectively, that were parked throughout the duration of the video. The images are part of the frames when the car was not completely obstructed by traffic.
The lilac and yellow colored components correspond to cars that show a specific profile.
Specifically, the yellow component consists of different cars that were captured by the camera and correspond to cars that stopped at the traffic light.

The red component has images of the same person positioned at the same place. Note that this person was present throughout the video. On the other hand, the orange component shows a greater diversity of people passing through the camera's view.
One can also see that the path from the red component (clear image of a person) to the green component (clear image of a car) is via the orange and lilac components, both of which correspond to more blurry images, which indicates that the model clearly differentiates between the images when the objects in them are clear.

\hide{
\begin{table}[h]
    \centering
    \caption{Number of classes inside each component according to YOLOv8.}
    \label{tab:YOLOv8-components}
    \begin{tabular}{|c|c|c|c|c|c|}
        \hline
        \textbf{Component} & \textbf{ \#Person} &  \textbf{ \#Bicycle} & \textbf{ \# Car}\\
        \hline
        Orange & 0 &0 & 30085 \\\hline
        Red & 0 &0 & 26050  \\\hline
        Green & 0 &0 & 9159  \\\hline
        Blue & 4505 &0 & 0  \\\hline
        Yellow & 18868 &0 & 0  \\\hline
        Lilac & 0 &0 & 5472  \\\hline
    \end{tabular}
\end{table}
}

\section{Limitations and Discussions}
\label{sec:limitations}

As seen in the previous Section, the addition of the nested TreeMap for exploration and the ability to highlight features of interest in TopoMap++ help users easily analyze high-dimensional datasets. At the same time, there are some limitations to these techniques that we aim to tackle going forward.

\myparagraph{Improving TreeMap Interaction}
Recall from Section~\ref{sec:treemap} that the rectangles for the TreeMap are created only when a component has at least $\eta$ points. Thus, a parent box is composed of points that are within its child boxes together with these outlier points.
One shortcoming of this is that it does not allow users to select such outliers. One way we are thinking of incorporating this is to add an explicitly marked ``outlier box" which can then be used to explore such points.

\myparagraph{Parameter identification}
Using TopoMap++ requires users to specify a few parameters.
First, there is $\eta$ which defines the simplification threshold, which is set $\eta$ to be approximately $1\%$ of the input size.
While this gave us good results for the data used in this paper, this might not always be the case.
Second is the parameter $c$, which is used to compute the scale factor $\alpha$. In this paper, we fix the value of $c$ to be 2. But, we could instead try to use the amount of white space present in the original TopoMap visualization 
to compute $c$ such that the resulting TopoMap++ visualization minimizes the amount of wasted space.

\section{Conclusion}
\label{sec:conclusion}

In this paper, we presented TopoMap++, 
that improves the visual space usage of the TopoMap~\cite{doraiswamy2020topomap} projection. This is accomplished by selectively scaling, and thus highlighting, topological features of interest. 
This reduces the space usage of the outliers that were the primary cause of the wasted space in the original TopoMap projection. While the resulting global layout no longer provides the topological guarantees of the original projection, these guarantees are still locally preserved within each highlighted component.
We also make use of the hierarchy formed by topological simplification to design a nested TreeMap-based visualization that allows users to easily interact with and analyze high-dimensional datasets using the TopoMap++ projection.
We also propose an approximation scheme to compute the Rips filtration inspired by state-of-the-art ANNS algorithms. We show that this approximation preserves the topology of the data with at least a two orders of magnitude improvement in the computational cost.

This work still focuses on the topology of the 0-cycles of the Rips filtration. However, given the complex topologies that are possible in higher dimensions, in the future, we intend to explore techniques that would enable TopoMap/TopoMap++ to also preserve/portray 1-cycles.

\acknowledgments{%
This work was supported by DARPA ASKEM, DARPA PTG, Capital One, and the NSF award CSSI 2411221. Any opinions, findings, conclusions, or recommendations expressed in this material are those of the authors and do not necessarily reflect the views of DARPA, Capital One, or NSF.
}

% \clearpage

\balance
\bibliographystyle{abbrv-doi-hyperref}

\bibliography{references}

\begin{thebibliography}{10}

\bibitem{ashraf2023survey}
M.~Ashraf, F.~Anowar, J.~H. Setu, A.~I. Chowdhury, E.~Ahmed, A.~Islam, and A.~Al-Mamun.
\newblock A survey on dimensionality reduction techniques for time-series data.
\newblock {\em IEEE Access}, 11:42909--42923, 2023. \href{https://doi.org/10.1109/ACCESS.2023.3269693}
{doi: {{%
10\hspace{.1pt}\discretionary{.}{%
}{.}\hspace{.4pt}1109\discretionary{/}{%
}{/}ACCESS\hspace{.1pt}\discretionary{.}{%
}{.}\hspace{.4pt}2023\hspace{.1pt}\discretionary{.}{%
}{.}\hspace{.4pt}3269693}}}


\bibitem{ayesha2020overview}
S.~Ayesha, M.~K. Hanif, and R.~Talib.
\newblock Overview and comparative study of dimensionality reduction techniques for high dimensional data.
\newblock {\em Information Fusion}, 59:44--58, 2020. \href{https://doi.org/10.1016/j.inffus.2020.01.005}
{doi: {{%
10\hspace{.1pt}\discretionary{.}{%
}{.}\hspace{.4pt}1016\discretionary{/}{%
}{/}j\hspace{.1pt}\discretionary{.}{%
}{.}\hspace{.4pt}inffus\hspace{.1pt}\discretionary{.}{%
}{.}\hspace{.4pt}2020\hspace{.1pt}\discretionary{.}{%
}{.}\hspace{.4pt}01\hspace{.1pt}\discretionary{.}{%
}{.}\hspace{.4pt}005}}}


\bibitem{bengio2004out}
Y.~Bengio, J.-f. Paiement, P.~Vincent, O.~Delalleau, N.~Roux, and M.~Ouimet.
\newblock Out-of-sample extensions for lle, isomap, mds, eigenmaps, and spectral clustering.
\newblock In S.~Thrun, L.~Saul, and B.~Sch\"{o}lkopf, eds., {\em Advances in Neural Information Processing Systems}, vol.~16. MIT Press, 2003.

\bibitem{Borg2006}
I.~Borg and P.~Groenen.
\newblock Modern multidimensional scaling: Theory and applications.
\newblock {\em Journal of Educational Measurement}, 40:277 -- 280, 06 2006. \href{https://doi.org/10.1111/j.1745-3984.2003.tb01108.x}
{doi: {{%
10\hspace{.1pt}\discretionary{.}{%
}{.}\hspace{.4pt}1111\discretionary{/}{%
}{/}j\hspace{.1pt}\discretionary{.}{%
}{.}\hspace{.4pt}1745\discretionary{%
}{-}{-}3984\hspace{.1pt}\discretionary{.}{%
}{.}\hspace{.4pt}2003\hspace{.1pt}\discretionary{.}{%
}{.}\hspace{.4pt}tb01108\hspace{.1pt}\discretionary{.}{%
}{.}\hspace{.4pt}x}}}


\bibitem{Carr2004Simplification}
H.~Carr, J.~Snoeyink, and M.~van~de Panne.
\newblock Simplifying flexible isosurfaces using local geometric measures.
\newblock In {\em IEEE Visualization 2004}, pp. 497--504, 2004. \href{https://doi.org/10.1109/VISUAL.2004.96}
{doi: {{%
10\hspace{.1pt}\discretionary{.}{%
}{.}\hspace{.4pt}1109\discretionary{/}{%
}{/}VISUAL\hspace{.1pt}\discretionary{.}{%
}{.}\hspace{.4pt}2004\hspace{.1pt}\discretionary{.}{%
}{.}\hspace{.4pt}96}}}


\bibitem{chao2019recent}
G.~Chao, Y.~Luo, and W.~Ding.
\newblock Recent advances in supervised dimension reduction: A survey.
\newblock {\em Machine Learning and Knowledge Extraction}, 1(1):341--358, 2019. \href{https://doi.org/10.3390/make1010020}
{doi: {{%
10\hspace{.1pt}\discretionary{.}{%
}{.}\hspace{.4pt}3390\discretionary{/}{%
}{/}make1010020}}}


\bibitem{chazal2021introduction}
F.~Chazal and B.~Michel.
\newblock An introduction to topological data analysis: Fundamental and practical aspects for data scientists.
\newblock {\em Frontiers in Artificial Intelligence}, 4, 2021. \href{https://doi.org/10.3389/frai.2021.667963}
{doi: {{%
10\hspace{.1pt}\discretionary{.}{%
}{.}\hspace{.4pt}3389\discretionary{/}{%
}{/}frai\hspace{.1pt}\discretionary{.}{%
}{.}\hspace{.4pt}2021\hspace{.1pt}\discretionary{.}{%
}{.}\hspace{.4pt}667963}}}


\bibitem{Clark2018ThinkYH}
P.~Clark, I.~Cowhey, O.~Etzioni, T.~Khot, A.~Sabharwal, C.~Schoenick, and O.~Tafjord.
\newblock Think you have solved question answering? try arc, the ai2 reasoning challenge, 2018.

\bibitem{cohen2005stability}
D.~Cohen-Steiner, H.~Edelsbrunner, and J.~Harer.
\newblock Stability of persistence diagrams.
\newblock In {\em Proceedings of the Twenty-First Annual Symposium on Computational Geometry}, SCG '05,  9 pages, p. 263–271. Association for Computing Machinery, New York, NY, USA, 2005. \href{https://doi.org/10.1145/1064092.1064133}
{doi: {{%
10\hspace{.1pt}\discretionary{.}{%
}{.}\hspace{.4pt}1145\discretionary{/}{%
}{/}1064092\hspace{.1pt}\discretionary{.}{%
}{.}\hspace{.4pt}1064133}}}


\bibitem{cunningham2015linear}
J.~P. Cunningham and Z.~Ghahramani.
\newblock Linear dimensionality reduction: Survey, insights, and generalizations.
\newblock {\em The Journal of Machine Learning Research}, 16(1):2859--2900, 2015.

\bibitem{mlpack2023}
R.~R. Curtin, M.~Edel, O.~Shrit, S.~Agrawal, S.~Basak, J.~J. Balamuta, R.~Birmingham, K.~Dutt, D.~Eddelbuettel, R.~Garg, S.~Jaiswal, A.~Kaushik, S.~Kim, A.~Mukherjee, N.~G. Sai, N.~Sharma, Y.~S. Parihar, R.~Swain, and C.~Sanderson.
\newblock mlpack 4: a fast, header-only c++ machine learning library.
\newblock {\em Journal of Open Source Software}, 8(82):5026, 2023. \href{https://doi.org/10.21105/joss.05026}
{doi: {{%
10\hspace{.1pt}\discretionary{.}{%
}{.}\hspace{.4pt}21105\discretionary{/}{%
}{/}joss\hspace{.1pt}\discretionary{.}{%
}{.}\hspace{.4pt}05026}}}


\bibitem{doppalapudi2022untangling}
B.~Doppalapudi, B.~Wang, and P.~Rosen.
\newblock Untangling force-directed layouts using persistent homology.
\newblock In {\em 2022 Topological Data Analysis and Visualization (TopoInVis)}, pp. 81--91, 2022. \href{https://doi.org/10.1109/TopoInVis57755.2022.00015}
{doi: {{%
10\hspace{.1pt}\discretionary{.}{%
}{.}\hspace{.4pt}1109\discretionary{/}{%
}{/}TopoInVis57755\hspace{.1pt}\discretionary{.}{%
}{.}\hspace{.4pt}2022\hspace{.1pt}\discretionary{.}{%
}{.}\hspace{.4pt}00015}}}


\bibitem{doraiswamy2020topomap}
H.~Doraiswamy, J.~Tierny, P.~J.~S. Silva, L.~G. Nonato, and C.~Silva.
\newblock Topomap: A 0-dimensional homology preserving projection of high-dimensional data.
\newblock {\em IEEE Transactions on Visualization and Computer Graphics}, 27(2):561--571, 2021. \href{https://doi.org/10.1109/TVCG.2020.3030441}
{doi: {{%
10\hspace{.1pt}\discretionary{.}{%
}{.}\hspace{.4pt}1109\discretionary{/}{%
}{/}TVCG\hspace{.1pt}\discretionary{.}{%
}{.}\hspace{.4pt}2020\hspace{.1pt}\discretionary{.}{%
}{.}\hspace{.4pt}3030441}}}


\bibitem{Dosovitskiy2020AnII}
A.~Dosovitskiy, L.~Beyer, A.~Kolesnikov, D.~Weissenborn, X.~Zhai, T.~Unterthiner, M.~Dehghani, M.~Minderer, G.~Heigold, S.~Gelly, J.~Uszkoreit, and N.~Houlsby.
\newblock An image is worth 16x16 words: Transformers for image recognition at scale.
\newblock In {\em International Conference on Learning Representations}, 2021.

\bibitem{Dua:2019UCI_datasets}
D.~Dua and C.~Graff.
\newblock {UCI} machine learning repository, 2017.

\bibitem{edelsbrunner2002topological}
Edelsbrunner, Letscher, and Zomorodian.
\newblock Topological persistence and simplification.
\newblock {\em Discrete \& computational geometry}, 28:511--533, 2002. \href{https://doi.org/10.1007/s00454-002-2885-2}
{doi: {{%
10\hspace{.1pt}\discretionary{.}{%
}{.}\hspace{.4pt}1007\discretionary{/}{%
}{/}s00454\discretionary{%
}{-}{-}002\discretionary{%
}{-}{-}2885\discretionary{%
}{-}{-}2}}}


\bibitem{DBLP:books/daglib/0025666}
H.~Edelsbrunner and J.~Harer.
\newblock {\em Computational Topology - an Introduction}.
\newblock American Mathematical Society, 2010.

\bibitem{engel2012survey}
D.~Engel, L.~H\"{u}ttenberger, and B.~Hamann.
\newblock {A Survey of Dimension Reduction Methods for High-dimensional Data Analysis and Visualization}.
\newblock In C.~Garth, A.~Middel, and H.~Hagen, eds., {\em Visualization of Large and Unstructured Data Sets: Applications in Geospatial Planning, Modeling and Engineering - Proceedings of IRTG 1131 Workshop 2011}, vol.~27 of {\em Open Access Series in Informatics (OASIcs)}, pp. 135--149. Schloss Dagstuhl -- Leibniz-Zentrum f{\"u}r Informatik, Dagstuhl, Germany, 2012. \href{https://doi.org/10.4230/OASIcs.VLUDS.2011.135}
{doi: {{%
10\hspace{.1pt}\discretionary{.}{%
}{.}\hspace{.4pt}4230\discretionary{/}{%
}{/}OASIcs\hspace{.1pt}\discretionary{.}{%
}{.}\hspace{.4pt}VLUDS\hspace{.1pt}\discretionary{.}{%
}{.}\hspace{.4pt}2011\hspace{.1pt}\discretionary{.}{%
}{.}\hspace{.4pt}135}}}


\bibitem{espadoto2019toward}
M.~Espadoto, R.~M. Martins, A.~Kerren, N.~S.~T. Hirata, and A.~C. Telea.
\newblock Toward a quantitative survey of dimension reduction techniques.
\newblock {\em IEEE Transactions on Visualization and Computer Graphics}, 27(3):2153--2173, 2021. \href{https://doi.org/10.1109/TVCG.2019.2944182}
{doi: {{%
10\hspace{.1pt}\discretionary{.}{%
}{.}\hspace{.4pt}1109\discretionary{/}{%
}{/}TVCG\hspace{.1pt}\discretionary{.}{%
}{.}\hspace{.4pt}2019\hspace{.1pt}\discretionary{.}{%
}{.}\hspace{.4pt}2944182}}}


\bibitem{gabriel69}
K.~R. Gabriel and R.~R. Sokal.
\newblock {A New Statistical Approach to Geographic Variation Analysis}.
\newblock {\em Systematic Biology}, 18(3):259--278, 09 1969. \href{https://doi.org/10.2307/2412323}
{doi: {{%
10\hspace{.1pt}\discretionary{.}{%
}{.}\hspace{.4pt}2307\discretionary{/}{%
}{/}2412323}}}


\bibitem{gerber2010}
S.~Gerber, P.-T. Bremer, V.~Pascucci, and R.~Whitaker.
\newblock Visual exploration of high dimensional scalar functions.
\newblock {\em IEEE Transactions on Visualization and Computer Graphics}, 16(6):1271--1280, 2010. \href{https://doi.org/10.1109/TVCG.2010.213}
{doi: {{%
10\hspace{.1pt}\discretionary{.}{%
}{.}\hspace{.4pt}1109\discretionary{/}{%
}{/}TVCG\hspace{.1pt}\discretionary{.}{%
}{.}\hspace{.4pt}2010\hspace{.1pt}\discretionary{.}{%
}{.}\hspace{.4pt}213}}}


\bibitem{gerber2013}
S.~Gerber, O.~Rübel, P.-T. Bremer, V.~Pascucci, and R.~T. Whitaker.
\newblock Morse–smale regression.
\newblock {\em Journal of Computational and Graphical Statistics}, 22(1):193--214, 2013.
\newblock PMID: 23687424. \href{https://doi.org/10.1080/10618600.2012.657132}
{doi: {{%
10\hspace{.1pt}\discretionary{.}{%
}{.}\hspace{.4pt}1080\discretionary{/}{%
}{/}10618600\hspace{.1pt}\discretionary{.}{%
}{.}\hspace{.4pt}2012\hspace{.1pt}\discretionary{.}{%
}{.}\hspace{.4pt}657132}}}


\bibitem{HWTopo10}
W.~Harvey and Y.~Wang.
\newblock Topological landscape ensembles for visualization of scalar-valued functions.
\newblock {\em Computer Graphics Forum}, 29(3):993--1002, 2010. \href{https://doi.org/10.1111/j.1467-8659.2009.01706.x}
{doi: {{%
10\hspace{.1pt}\discretionary{.}{%
}{.}\hspace{.4pt}1111\discretionary{/}{%
}{/}j\hspace{.1pt}\discretionary{.}{%
}{.}\hspace{.4pt}1467\discretionary{%
}{-}{-}8659\hspace{.1pt}\discretionary{.}{%
}{.}\hspace{.4pt}2009\hspace{.1pt}\discretionary{.}{%
}{.}\hspace{.4pt}01706\hspace{.1pt}\discretionary{.}{%
}{.}\hspace{.4pt}x}}}


\bibitem{geoffrey:2002:nips}
G.~E. Hinton and S.~Roweis.
\newblock Stochastic neighbor embedding.
\newblock In S.~Becker, S.~Thrun, and K.~Obermayer, eds., {\em Advances in Neural Information Processing Systems}, vol.~15. MIT Press, 2002.

\bibitem{NEURIPS2019_Vamana}
S.~Jayaram~Subramanya, F.~Devvrit, H.~V. Simhadri, R.~Krishnawamy, and R.~Kadekodi.
\newblock Diskann: Fast accurate billion-point nearest neighbor search on a single node.
\newblock In H.~Wallach, H.~Larochelle, A.~Beygelzimer, F.~d\textquotesingle Alch\'{e}-Buc, E.~Fox, and R.~Garnett, eds., {\em Advances in Neural Information Processing Systems}, vol.~32. Curran Associates, Inc., 2019.

\bibitem{jenkins2004spatio}
O.~C. Jenkins and M.~J. Matari\'{c}.
\newblock A spatio-temporal extension to isomap nonlinear dimension reduction.
\newblock In {\em Proceedings of the Twenty-First International Conference on Machine Learning}, ICML '04, p.~56. Association for Computing Machinery, New York, NY, USA, 2004. \href{https://doi.org/10.1145/1015330.1015357}
{doi: {{%
10\hspace{.1pt}\discretionary{.}{%
}{.}\hspace{.4pt}1145\discretionary{/}{%
}{/}1015330\hspace{.1pt}\discretionary{.}{%
}{.}\hspace{.4pt}1015357}}}


\bibitem{yolov8_ultralytics}
G.~Jocher, A.~Chaurasia, and J.~Qiu.
\newblock Ultralytics yolov8, 2023.

\bibitem{BIGANN_DATASET}
H.~Jégou, R.~Tavenard, M.~Douze, and L.~Amsaleg.
\newblock Searching in one billion vectors: Re-rank with source coding.
\newblock In {\em 2011 IEEE International Conference on Acoustics, Speech and Signal Processing (ICASSP)}, pp. 861--864, 2011. \href{https://doi.org/10.1109/ICASSP.2011.5946540}
{doi: {{%
10\hspace{.1pt}\discretionary{.}{%
}{.}\hspace{.4pt}1109\discretionary{/}{%
}{/}ICASSP\hspace{.1pt}\discretionary{.}{%
}{.}\hspace{.4pt}2011\hspace{.1pt}\discretionary{.}{%
}{.}\hspace{.4pt}5946540}}}


\bibitem{mnist}
Y.~LeCun and C.~Cortes.
\newblock {MNIST} handwritten digit database.
\newblock 2010.

\bibitem{lee2005nonlinear}
J.~A. Lee and M.~Verleysen.
\newblock Nonlinear dimensionality reduction of data manifolds with essential loops.
\newblock {\em Neurocomputing}, 67:29--53, 2005.
\newblock Geometrical Methods in Neural Networks and Learning. \href{https://doi.org/10.1016/j.neucom.2004.11.042}
{doi: {{%
10\hspace{.1pt}\discretionary{.}{%
}{.}\hspace{.4pt}1016\discretionary{/}{%
}{/}j\hspace{.1pt}\discretionary{.}{%
}{.}\hspace{.4pt}neucom\hspace{.1pt}\discretionary{.}{%
}{.}\hspace{.4pt}2004\hspace{.1pt}\discretionary{.}{%
}{.}\hspace{.4pt}11\hspace{.1pt}\discretionary{.}{%
}{.}\hspace{.4pt}042}}}


\bibitem{10.1145/1835804.1835882/FastEMST}
W.~B. March, P.~Ram, and A.~G. Gray.
\newblock Fast euclidean minimum spanning tree: algorithm, analysis, and applications.
\newblock In {\em Proceedings of the 16th ACM SIGKDD International Conference on Knowledge Discovery and Data Mining}, KDD '10,  10 pages, p. 603–612. Association for Computing Machinery, New York, NY, USA, 2010. \href{https://doi.org/10.1145/1835804.1835882}
{doi: {{%
10\hspace{.1pt}\discretionary{.}{%
}{.}\hspace{.4pt}1145\discretionary{/}{%
}{/}1835804\hspace{.1pt}\discretionary{.}{%
}{.}\hspace{.4pt}1835882}}}


\bibitem{mcinnes2018umap}
L.~McInnes, J.~Healy, N.~Saul, and L.~Großberger.
\newblock Umap: Uniform manifold approximation and projection.
\newblock {\em Journal of Open Source Software}, 3(29):861, 2018. \href{https://doi.org/10.21105/joss.00861}
{doi: {{%
10\hspace{.1pt}\discretionary{.}{%
}{.}\hspace{.4pt}21105\discretionary{/}{%
}{/}joss\hspace{.1pt}\discretionary{.}{%
}{.}\hspace{.4pt}00861}}}


\bibitem{nelson2022topology}
B.~J. Nelson and Y.~Luo.
\newblock Topology-preserving dimensionality reduction via interleaving optimization.
\newblock {\em CoRR}, abs/2201.13012, 2022.

\bibitem{nonato2018multidimensional}
L.~G. Nonato and M.~Aupetit.
\newblock Multidimensional projection for visual analytics: Linking techniques with distortions, tasks, and layout enrichment.
\newblock {\em IEEE Transactions on Visualization and Computer Graphics}, 25(8):2650--2673, 2019. \href{https://doi.org/10.1109/TVCG.2018.2846735}
{doi: {{%
10\hspace{.1pt}\discretionary{.}{%
}{.}\hspace{.4pt}1109\discretionary{/}{%
}{/}TVCG\hspace{.1pt}\discretionary{.}{%
}{.}\hspace{.4pt}2018\hspace{.1pt}\discretionary{.}{%
}{.}\hspace{.4pt}2846735}}}


\bibitem{OHJSH11}
P.~Oesterling, C.~Heine, H.~Janicke, G.~Scheuermann, and G.~Heyer.
\newblock Visualization of high-dimensional point clouds using their density distribution's topology.
\newblock {\em IEEE Transactions on Visualization and Computer Graphics}, 17(11):1547--1559, 2011. \href{https://doi.org/10.1109/TVCG.2011.27}
{doi: {{%
10\hspace{.1pt}\discretionary{.}{%
}{.}\hspace{.4pt}1109\discretionary{/}{%
}{/}TVCG\hspace{.1pt}\discretionary{.}{%
}{.}\hspace{.4pt}2011\hspace{.1pt}\discretionary{.}{%
}{.}\hspace{.4pt}27}}}


\bibitem{OesterlingHJS10}
P.~Oesterling, C.~Heine, H.~Jänicke, and G.~Scheuermann.
\newblock Visual analysis of high dimensional point clouds using topological landscapes.
\newblock In {\em 2010 IEEE Pacific Visualization Symposium (PacificVis)}, pp. 113--120, 2010. \href{https://doi.org/10.1109/PACIFICVIS.2010.5429601}
{doi: {{%
10\hspace{.1pt}\discretionary{.}{%
}{.}\hspace{.4pt}1109\discretionary{/}{%
}{/}PACIFICVIS\hspace{.1pt}\discretionary{.}{%
}{.}\hspace{.4pt}2010\hspace{.1pt}\discretionary{.}{%
}{.}\hspace{.4pt}5429601}}}


\bibitem{Oesterling0WS13}
P.~Oesterling, C.~Heine, G.~H. Weber, and G.~Scheuermann.
\newblock Visualizing nd point clouds as topological landscape profiles to guide local data analysis.
\newblock {\em IEEE Transactions on Visualization and Computer Graphics}, 19(3):514--526, 2013. \href{https://doi.org/10.1109/TVCG.2012.120}
{doi: {{%
10\hspace{.1pt}\discretionary{.}{%
}{.}\hspace{.4pt}1109\discretionary{/}{%
}{/}TVCG\hspace{.1pt}\discretionary{.}{%
}{.}\hspace{.4pt}2012\hspace{.1pt}\discretionary{.}{%
}{.}\hspace{.4pt}120}}}


\bibitem{OesterlingSTHKEW10}
P.~Oesterling, G.~Scheuermann, S.~Teresniak, G.~Heyer, S.~Koch, T.~Ertl, and G.~H. Weber.
\newblock Two-stage framework for a topology-based projection and visualization of classified document collections.
\newblock In {\em 2010 IEEE Symposium on Visual Analytics Science and Technology}, pp. 91--98, 2010. \href{https://doi.org/10.1109/VAST.2010.5652940}
{doi: {{%
10\hspace{.1pt}\discretionary{.}{%
}{.}\hspace{.4pt}1109\discretionary{/}{%
}{/}VAST\hspace{.1pt}\discretionary{.}{%
}{.}\hspace{.4pt}2010\hspace{.1pt}\discretionary{.}{%
}{.}\hspace{.4pt}5652940}}}


\bibitem{piadyk_streetaware_2023}
Y.~Piadyk, J.~Rulff, E.~Brewer, M.~Hosseini, K.~Ozbay, M.~Sankaradas, S.~Chakradhar, and C.~Silva.
\newblock Streetaware: A high-resolution synchronized multimodal urban scene dataset.
\newblock {\em Sensors}, 23(7), 2023. \href{https://doi.org/10.3390/s23073710}
{doi: {{%
10\hspace{.1pt}\discretionary{.}{%
}{.}\hspace{.4pt}3390\discretionary{/}{%
}{/}s23073710}}}


\bibitem{ray2021various}
P.~Ray, S.~S. Reddy, and T.~Banerjee.
\newblock Various dimension reduction techniques for high dimensional data analysis: a review.
\newblock {\em Artificial Intelligence Review}, 54(5):3473--3515, June 2021. \href{https://doi.org/10.1007/s10462-020-09928-0}
{doi: {{%
10\hspace{.1pt}\discretionary{.}{%
}{.}\hspace{.4pt}1007\discretionary{/}{%
}{/}s10462\discretionary{%
}{-}{-}020\discretionary{%
}{-}{-}09928\discretionary{%
}{-}{-}0}}}


\bibitem{rieck2015persistent}
B.~Rieck and H.~Leitte.
\newblock Persistent homology for the evaluation of dimensionality reduction schemes.
\newblock {\em Computer Graphics Forum}, 34(3):431--440, 2015. \href{https://doi.org/10.1111/cgf.12655}
{doi: {{%
10\hspace{.1pt}\discretionary{.}{%
}{.}\hspace{.4pt}1111\discretionary{/}{%
}{/}cgf\hspace{.1pt}\discretionary{.}{%
}{.}\hspace{.4pt}12655}}}


\bibitem{rieck2017agreement}
B.~Rieck and H.~Leitte.
\newblock Agreement analysis of quality measures for dimensionality reduction.
\newblock In H.~Carr, C.~Garth, and T.~Weinkauf, eds., {\em Topological Methods in Data Analysis and Visualization IV}, pp. 103--117. Springer International Publishing, Cham, 2017.

\bibitem{sacha2016visual}
D.~Sacha, L.~Zhang, M.~Sedlmair, J.~A. Lee, J.~Peltonen, D.~Weiskopf, S.~C. North, and D.~A. Keim.
\newblock Visual interaction with dimensionality reduction: A structured literature analysis.
\newblock {\em IEEE Transactions on Visualization and Computer Graphics}, 23(1):241--250, 2017. \href{https://doi.org/10.1109/TVCG.2016.2598495}
{doi: {{%
10\hspace{.1pt}\discretionary{.}{%
}{.}\hspace{.4pt}1109\discretionary{/}{%
}{/}TVCG\hspace{.1pt}\discretionary{.}{%
}{.}\hspace{.4pt}2016\hspace{.1pt}\discretionary{.}{%
}{.}\hspace{.4pt}2598495}}}


\bibitem{sedlmair2013empirical}
M.~Sedlmair, T.~Munzner, and M.~Tory.
\newblock Empirical guidance on scatterplot and dimension reduction technique choices.
\newblock {\em IEEE Transactions on Visualization and Computer Graphics}, 19(12):2634--2643, 2013. \href{https://doi.org/10.1109/TVCG.2013.153}
{doi: {{%
10\hspace{.1pt}\discretionary{.}{%
}{.}\hspace{.4pt}1109\discretionary{/}{%
}{/}TVCG\hspace{.1pt}\discretionary{.}{%
}{.}\hspace{.4pt}2013\hspace{.1pt}\discretionary{.}{%
}{.}\hspace{.4pt}153}}}


\bibitem{shneiderman1992tree}
B.~Shneiderman.
\newblock Tree visualization with tree-maps: 2-d space-filling approach.
\newblock {\em ACM Trans. Graph.}, 11(1):92–99,  8 pages, jan 1992. \href{https://doi.org/10.1145/102377.115768}
{doi: {{%
10\hspace{.1pt}\discretionary{.}{%
}{.}\hspace{.4pt}1145\discretionary{/}{%
}{/}102377\hspace{.1pt}\discretionary{.}{%
}{.}\hspace{.4pt}115768}}}


\bibitem{silva2003global}
V.~Silva and J.~Tenenbaum.
\newblock Global versus local methods in nonlinear dimensionality reduction.
\newblock In S.~Becker, S.~Thrun, and K.~Obermayer, eds., {\em Advances in Neural Information Processing Systems}, vol.~15. MIT Press, 2002.

\bibitem{sohns2021attribute}
J.-T. Sohns, M.~Schmitt, F.~Jirasek, H.~Hasse, and H.~Leitte.
\newblock Attribute-based explanation of non-linear embeddings of high-dimensional data.
\newblock {\em IEEE Transactions on Visualization and Computer Graphics}, 28(1):540--550, 2022. \href{https://doi.org/10.1109/TVCG.2021.3114870}
{doi: {{%
10\hspace{.1pt}\discretionary{.}{%
}{.}\hspace{.4pt}1109\discretionary{/}{%
}{/}TVCG\hspace{.1pt}\discretionary{.}{%
}{.}\hspace{.4pt}2021\hspace{.1pt}\discretionary{.}{%
}{.}\hspace{.4pt}3114870}}}


\bibitem{tenenbaum2000global}
J.~B. Tenenbaum, V.~de~Silva, and J.~C. Langford.
\newblock A global geometric framework for nonlinear dimensionality reduction.
\newblock {\em Science}, 290(5500):2319--2323, 2000. \href{https://doi.org/10.1126/science.290.5500.2319}
{doi: {{%
10\hspace{.1pt}\discretionary{.}{%
}{.}\hspace{.4pt}1126\discretionary{/}{%
}{/}science\hspace{.1pt}\discretionary{.}{%
}{.}\hspace{.4pt}290\hspace{.1pt}\discretionary{.}{%
}{.}\hspace{.4pt}5500\hspace{.1pt}\discretionary{.}{%
}{.}\hspace{.4pt}2319}}}


\bibitem{touvron2023llama}
H.~Touvron, L.~Martin, K.~Stone, P.~Albert, A.~Almahairi, Y.~Babaei, N.~Bashlykov, S.~Batra, P.~Bhargava, S.~Bhosale, D.~Bikel, L.~Blecher, C.~Canton-Ferrer, M.~Chen, G.~Cucurull, D.~Esiobu, J.~Fernandes, J.~Fu, W.~Fu, B.~Fuller, C.~Gao, V.~Goswami, N.~Goyal, A.~Hartshorn, S.~Hosseini, R.~Hou, H.~Inan, M.~Kardas, V.~Kerkez, M.~Khabsa, I.~Kloumann, A.~Korenev, P.~S. Koura, M.-A. Lachaux, T.~Lavril, J.~Lee, D.~Liskovich, Y.~Lu, Y.~Mao, X.~Martinet, T.~Mihaylov, P.~Mishra, I.~Molybog, Y.~Nie, A.~Poulton, J.~Reizenstein, R.~Rungta, K.~Saladi, A.~Schelten, R.~Silva, E.~M. Smith, R.~Subramanian, X.~E. Tan, B.~Tang, R.~Taylor, A.~Williams, J.~X. Kuan, P.~Xu, Z.~Yan, I.~Zarov, Y.~Zhang, A.~Fan, M.~Kambadur, S.~Narang, A.~Rodriguez, R.~Stojnic, S.~Edunov, and T.~Scialom.
\newblock Llama 2: Open foundation and fine-tuned chat models.
\newblock {\em CoRR}, abs/2307.09288, 2023.

\bibitem{van2008visualizing}
L.~van~der Maaten and G.~Hinton.
\newblock Visualizing data using t-sne.
\newblock {\em Journal of Machine Learning Research}, 9(86):2579--2605, 2008.

\bibitem{Maaten2007review}
L.~van~der Maaten, E.~Postma, and H.~Herik.
\newblock Dimensionality reduction: A comparative review.
\newblock {\em Journal of Machine Learning Research - JMLR}, 10, 01 2007.

\bibitem{vandaele2021topologically}
R.~Vandaele, B.~Kang, J.~Lijffijt, T.~D. Bie, and Y.~Saeys.
\newblock Topologically regularized data embeddings.
\newblock In {\em International Conference on Learning Representations}, 2022.

\bibitem{velliangiri2019review}
S.~Velliangiri, S.~Alagumuthukrishnan, and S.~I. {Thankumar joseph}.
\newblock A review of dimensionality reduction techniques for efficient computation.
\newblock {\em Procedia Computer Science}, 165:104--111, 2019.
\newblock 2nd International Conference on Recent Trends in Advanced Computing ICRTAC -DISRUP - TIV INNOVATION , 2019 November 11-12, 2019. \href{https://doi.org/10.1016/j.procs.2020.01.079}
{doi: {{%
10\hspace{.1pt}\discretionary{.}{%
}{.}\hspace{.4pt}1016\discretionary{/}{%
}{/}j\hspace{.1pt}\discretionary{.}{%
}{.}\hspace{.4pt}procs\hspace{.1pt}\discretionary{.}{%
}{.}\hspace{.4pt}2020\hspace{.1pt}\discretionary{.}{%
}{.}\hspace{.4pt}01\hspace{.1pt}\discretionary{.}{%
}{.}\hspace{.4pt}079}}}


\bibitem{vu2022integrating}
V.~M. Vu, A.~Bibal, and B.~Frénay.
\newblock Integrating constraints into dimensionality reduction for visualization: A survey.
\newblock {\em IEEE Transactions on Artificial Intelligence}, 3(6):944--962, 2022. \href{https://doi.org/10.1109/TAI.2022.3204734}
{doi: {{%
10\hspace{.1pt}\discretionary{.}{%
}{.}\hspace{.4pt}1109\discretionary{/}{%
}{/}TAI\hspace{.1pt}\discretionary{.}{%
}{.}\hspace{.4pt}2022\hspace{.1pt}\discretionary{.}{%
}{.}\hspace{.4pt}3204734}}}


\bibitem{wagner2021improving}
A.~Wagner, E.~Solomon, and P.~Bendich.
\newblock Improving metric dimensionality reduction with distributed topology.
\newblock {\em CoRR}, abs/2106.07613, 2021.

\bibitem{Weber:2007}
G.~Weber, P.-T. Bremer, and V.~Pascucci.
\newblock Topological landscapes: A terrain metaphor for scientific data.
\newblock {\em IEEE Transactions on Visualization and Computer Graphics}, 13(6):1416--1423, 2007. \href{https://doi.org/10.1109/TVCG.2007.70601}
{doi: {{%
10\hspace{.1pt}\discretionary{.}{%
}{.}\hspace{.4pt}1109\discretionary{/}{%
}{/}TVCG\hspace{.1pt}\discretionary{.}{%
}{.}\hspace{.4pt}2007\hspace{.1pt}\discretionary{.}{%
}{.}\hspace{.4pt}70601}}}


\bibitem{xia2021revisiting}
J.~Xia, Y.~Zhang, J.~Song, Y.~Chen, Y.~Wang, and S.~Liu.
\newblock Revisiting dimensionality reduction techniques for visual cluster analysis: An empirical study.
\newblock {\em IEEE Transactions on Visualization and Computer Graphics}, 28(1):529--539, 2022. \href{https://doi.org/10.1109/TVCG.2021.3114694}
{doi: {{%
10\hspace{.1pt}\discretionary{.}{%
}{.}\hspace{.4pt}1109\discretionary{/}{%
}{/}TVCG\hspace{.1pt}\discretionary{.}{%
}{.}\hspace{.4pt}2021\hspace{.1pt}\discretionary{.}{%
}{.}\hspace{.4pt}3114694}}}


\bibitem{yan2018homology}
L.~Yan, Y.~Zhao, P.~Rosen, C.~Scheidegger, and B.~Wang.
\newblock Homology-preserving dimensionality reduction via manifold landmarking and tearing.
\newblock {\em CoRR}, abs/1806.08460, 2018.

\end{thebibliography}

\end{document}

%%%%%%%%%%%%%%%%%%%%%%%%%%%%%%%%%%%%%%%%%%%%%%%%%%%%%%%%%%%%%%%%
%%%%%%%%%%%%%%%%%%%%%% START OF THE PAPER %%%%%%%%%%%%%%%%%%%%%%
%%%%%%%%%%%%%%%%%%%%%%%%%%%%%%%%%%%%%%%%%%%%%%%%%%%%%%%%%%%%%%%%


\maketitle

\appendix

\section{Quantitative Evaluation}

In this section, we describe additional experiments that further evaluate the approximation power of the proposed approach that computes the Euclidean minimum spanning tree.
EMST computation is expensive, especially when the dimension of the data is large.
Since we are only interested in performing an approximation for such data, we exclude the smaller dimension datasets from these experiments.

\subsection{Sensitivity to Parameters}

We first look at the sensitivity of the approximation scheme with respect to the different parameters used.

\noindent\textbf{1.~Parameter $\alpha$: }
According to the official documentation~\cite{diskann-github}, it is recommended to have an $\alpha$ value between 1.0 and 1.5. This parameter directly impacts the diameter of the generated graph, which is approximately $\log_\alpha n$. Therefore, it is expected that the higher the $\alpha$, the longer it takes to construct the AMST and the better the quality of the approximation.

\begin{figure}[!b]
    \centering
    \includegraphics[width=\linewidth]{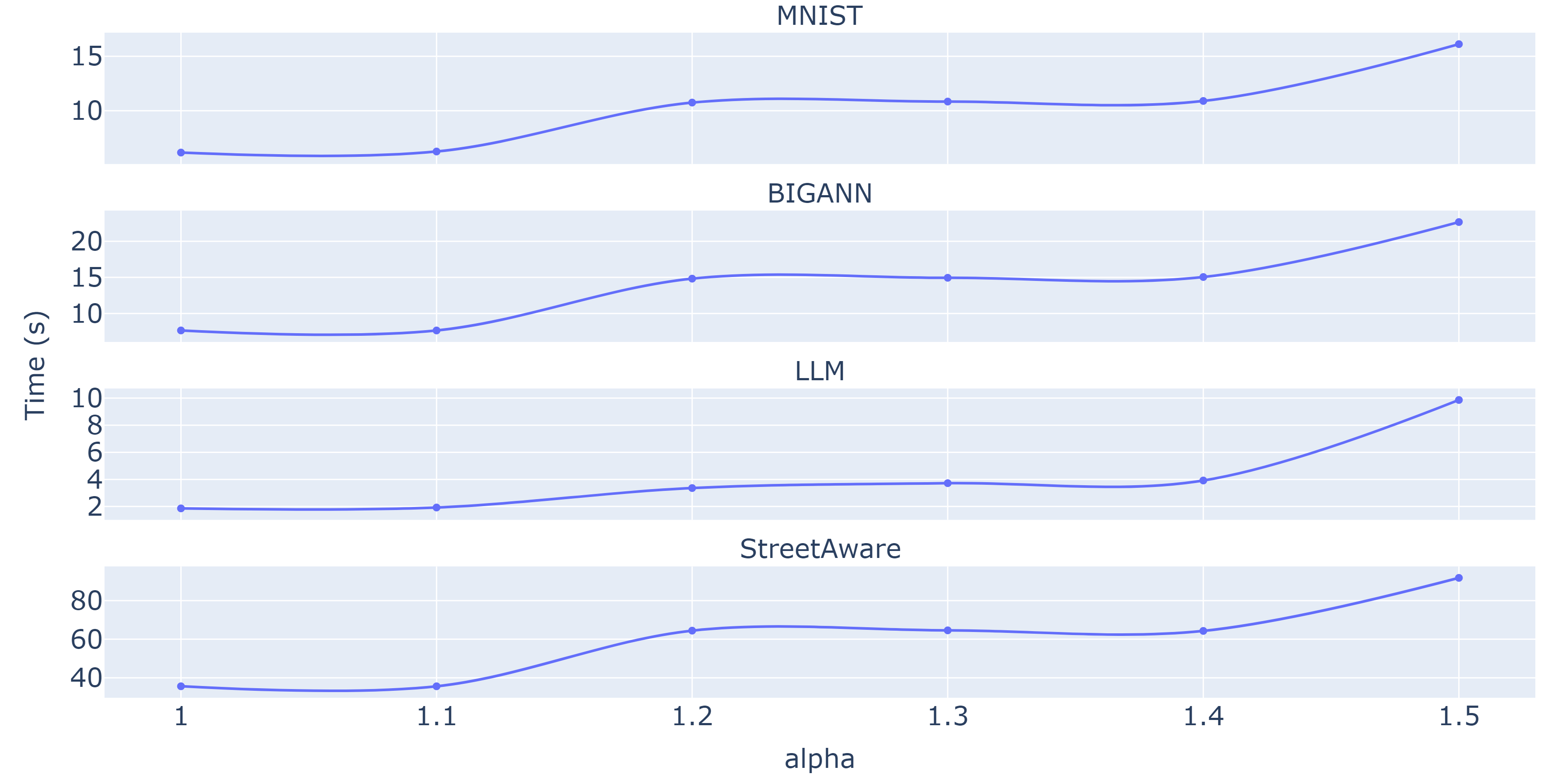}
    \caption{Time to compute the AMST with varying $\alpha$.}
    \label{fig:impact-alpha-time}
    
    \centering
    \includegraphics[width=\linewidth]{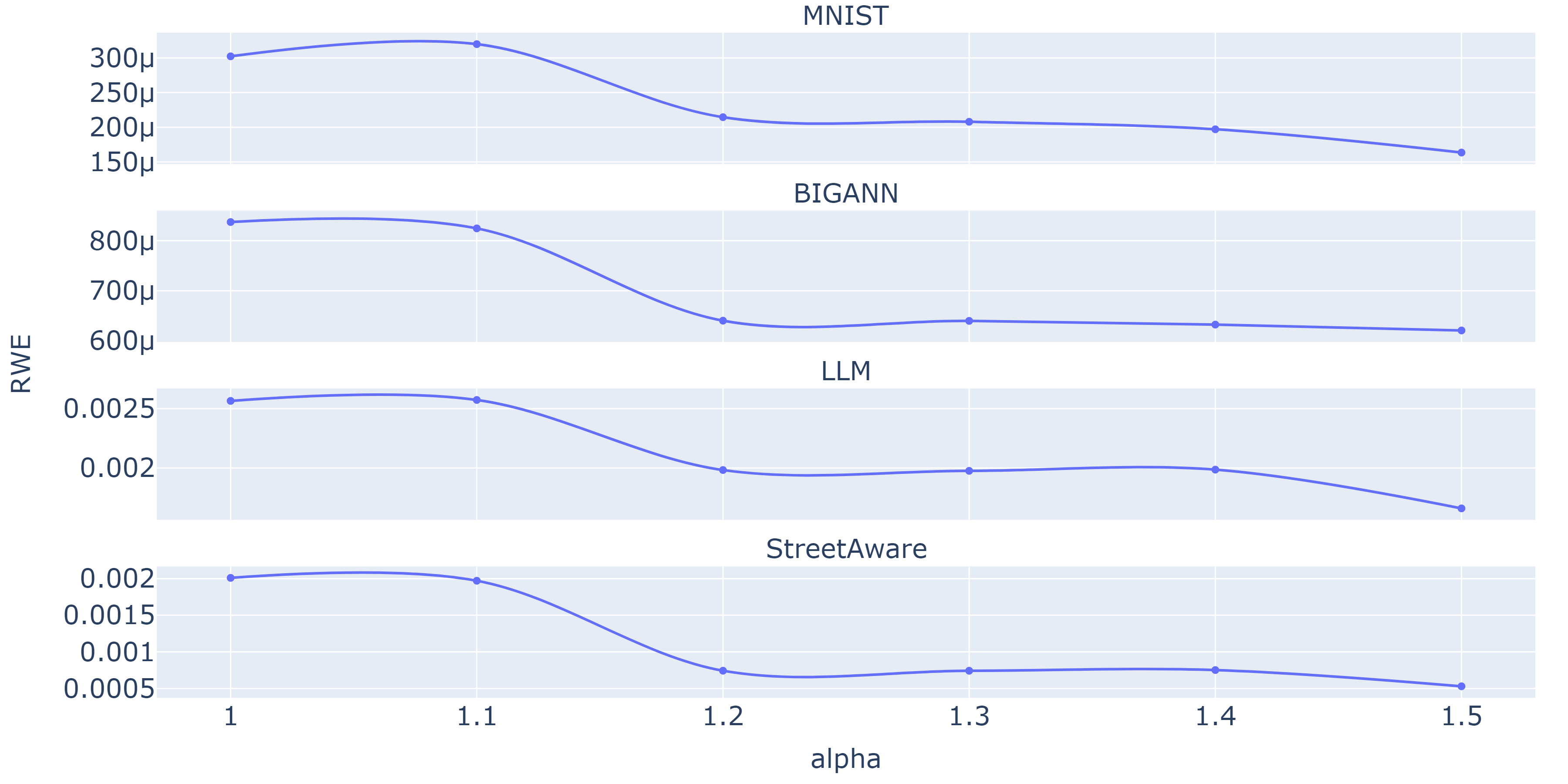}
    \vspace{-0.2in}
    \caption{RWE between the AMST and EMST with varying $\alpha$.}
    \label{fig:impact-alpha-RWE}

    \centering
    \includegraphics[width=\linewidth]{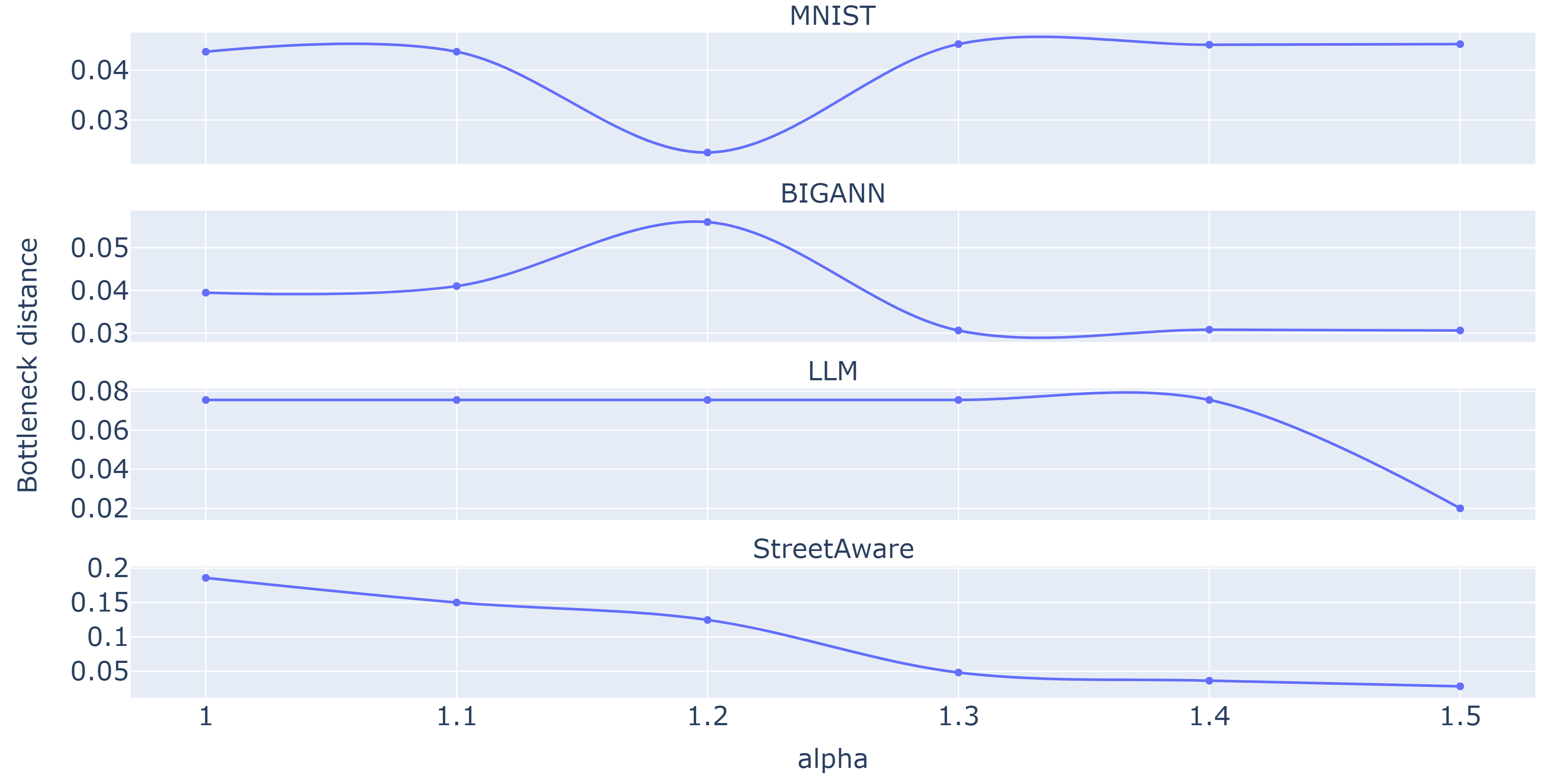}
    \vspace{-0.2in}
    \caption{Bottleneck distance with varying $\alpha$.}
    \label{fig:impact-alpha-bottleneck}
\end{figure}

To analyze the impact of this parameter, we set $R=L=100$ and varied $\alpha$ between 1 and 1.5. As we can see in Figure~\ref{fig:impact-alpha-time}, the time needed to build the AMST increases with increasing alpha value.
% but all times remain in the same order of magnitude as the original results. 
Furthermore, Figure~\ref{fig:impact-alpha-RWE} shows how this parameter influences $RWE$, where we see that errors tend to decrease with increasing $\alpha$. There is however some small fluctuation in the case of Bottleneck distance, as seen in Figure~\ref{fig:impact-alpha-bottleneck}.
% However, note that this fluctuation is quite small.
This could be due to the fact that the bottleneck distance looks at the maximum deviation between matching points within the persistence diagram. Thus, even a single edge in the approximate MST with a small weight difference can easily affect this metric.

\noindent\textbf{2.~Parameter $L$: }
This parameter defines the size of the search list maintained during the Vamana Index construction. The recommended values range from 75 to 200, with a minimum recommended value equal to that of $R$. Similar to $\alpha$, higher values of $L$ are expected to yield better results but require more time for building the index. For this analysis, we fixed $R=100$ and $\alpha = 1.3$ and varied $L$ from 100 to 200.

\begin{figure}[t]
    \centering
    \includegraphics[width=\linewidth]{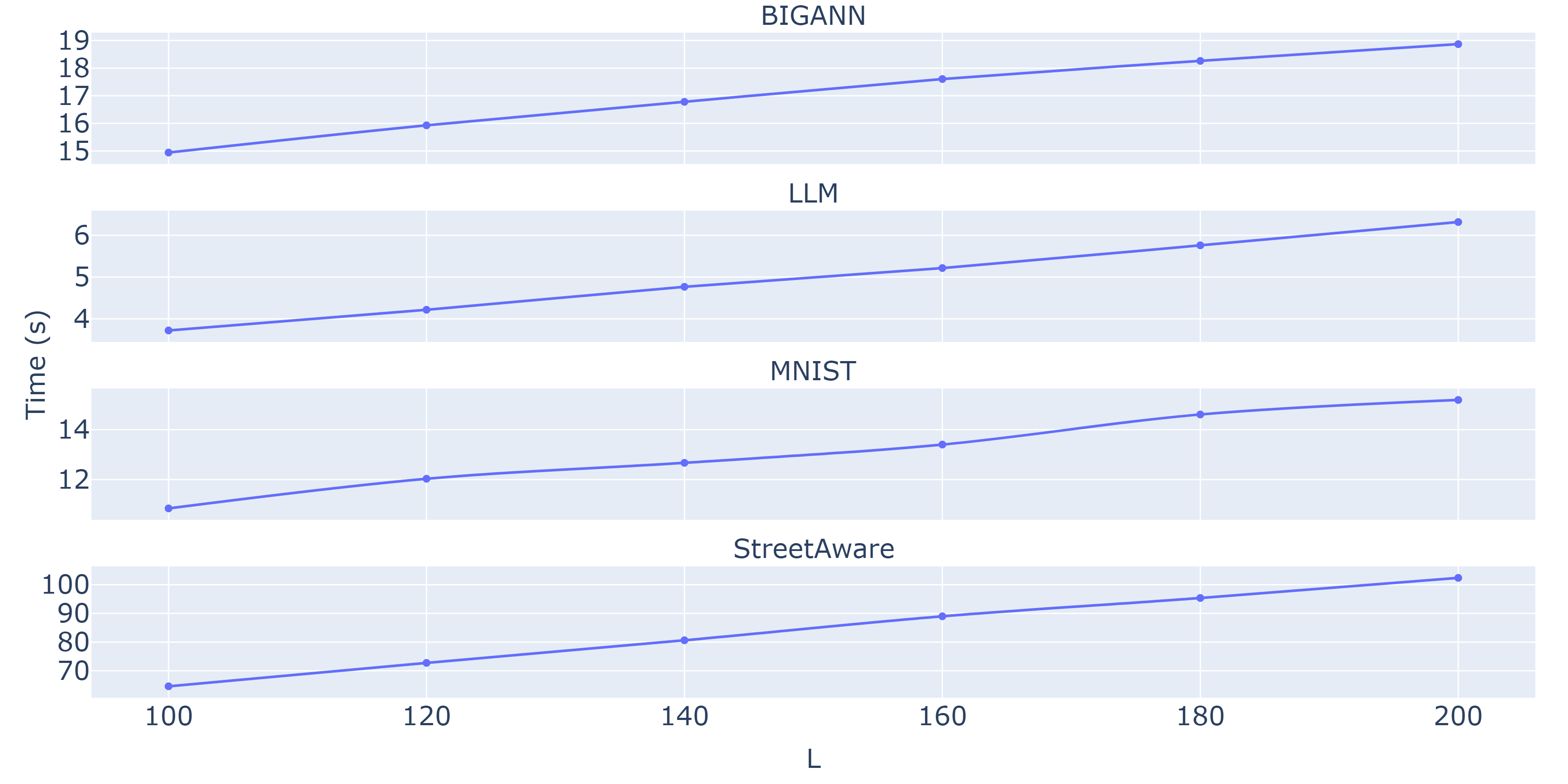}
    \caption{Time to compute the AMST with varying $L$.}
    \label{fig:impact-L-time}

    \centering
    \includegraphics[width=\linewidth]{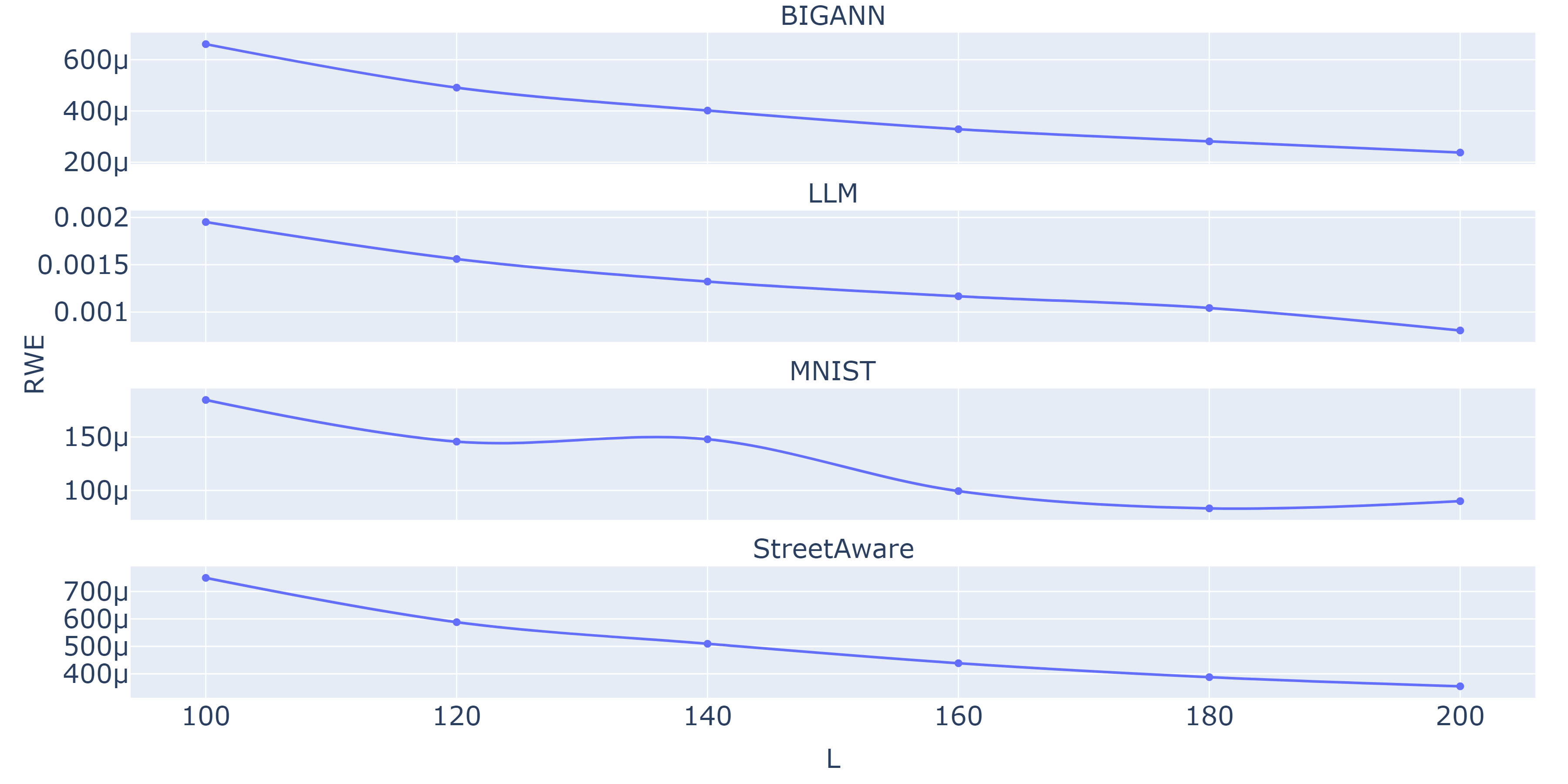}
    \caption{RWE between the AMST and EMST with varying $L$.}
    \label{fig:impact-L-RWE}

    \centering
    \includegraphics[width=\linewidth]{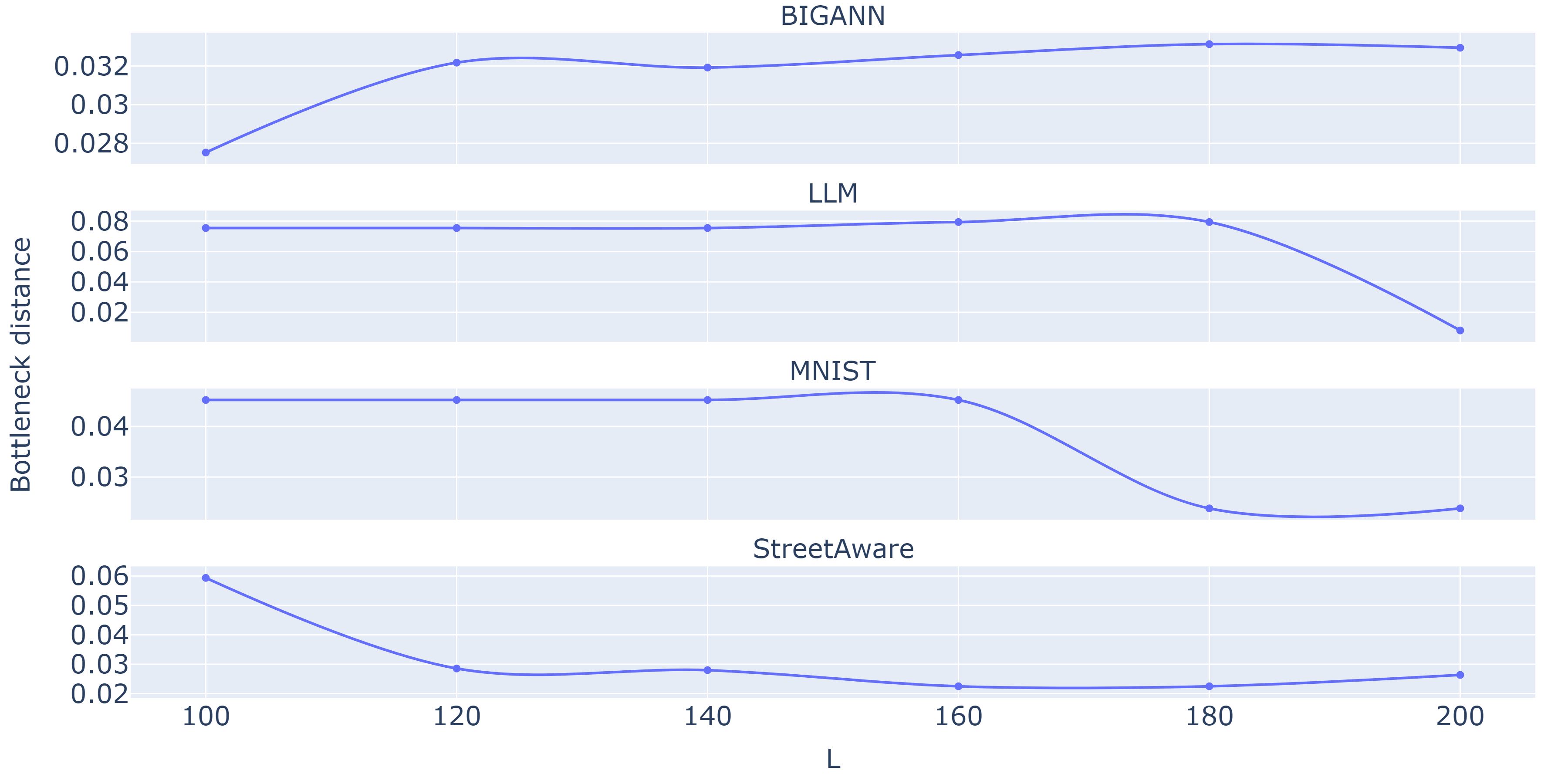}
    \caption{Bottleneck distance with varying $L$.}
    \label{fig:impact-L-bottleneck}
\end{figure}

Figure~\ref{fig:impact-L-time} shows how increasing $L$ impacts the time needed to compute the AMST. As expected, the running time increases with the increase in $L$.
% In all datasets the time either remained basically the same, as in the case of Seeds and Iris, or increased but maintained the same order of magnitude, as in the case of the largest datasets.
%
Figure \ref{fig:impact-L-RWE} shows how sensitive $RWE$ is with respect to $L$. We can see that the values tend to decrease or remain constant.
Finally, Figure \ref{fig:impact-L-bottleneck} shows the variation in bottleneck distance with increasing $L$. Again, the bottleneck distance tends to decrease with increasing $L$ except in the case of the BIGANN data, where we see that it increases, even though the RWE between the MSTs decreases. 
However, note that this increase is minute (the difference is in the order of $10^{-3}$).

% \begin{figure}[h]
%     \centering
%     \includegraphics[width=\linewidth]{figs/supp/time-alpha.png}
%     \caption{Impact of increasing the $\alpha$ parameter on time to compute the AMST}
%     \label{fig:impact-alpha-time}
    
%     \centering
%     \includegraphics[width=\linewidth]{figs/supp/RWE-alpha.png}
%     \caption{Impact of increasing the $\alpha$ parameter on the Relative Weigth error between the AMST and EMST}
%     \label{fig:impact-alpha-RWE}

%     \centering
%     \includegraphics[width=\linewidth]{figs/supp/BD-alpha.png}
%     \caption{Impact of increasing the $\alpha$ parameter on the Bottleneck distance}
%     \label{fig:impact-alpha-bottleneck}
% \end{figure}

\begin{figure}[t]
    \centering
    \includegraphics[width=\linewidth]{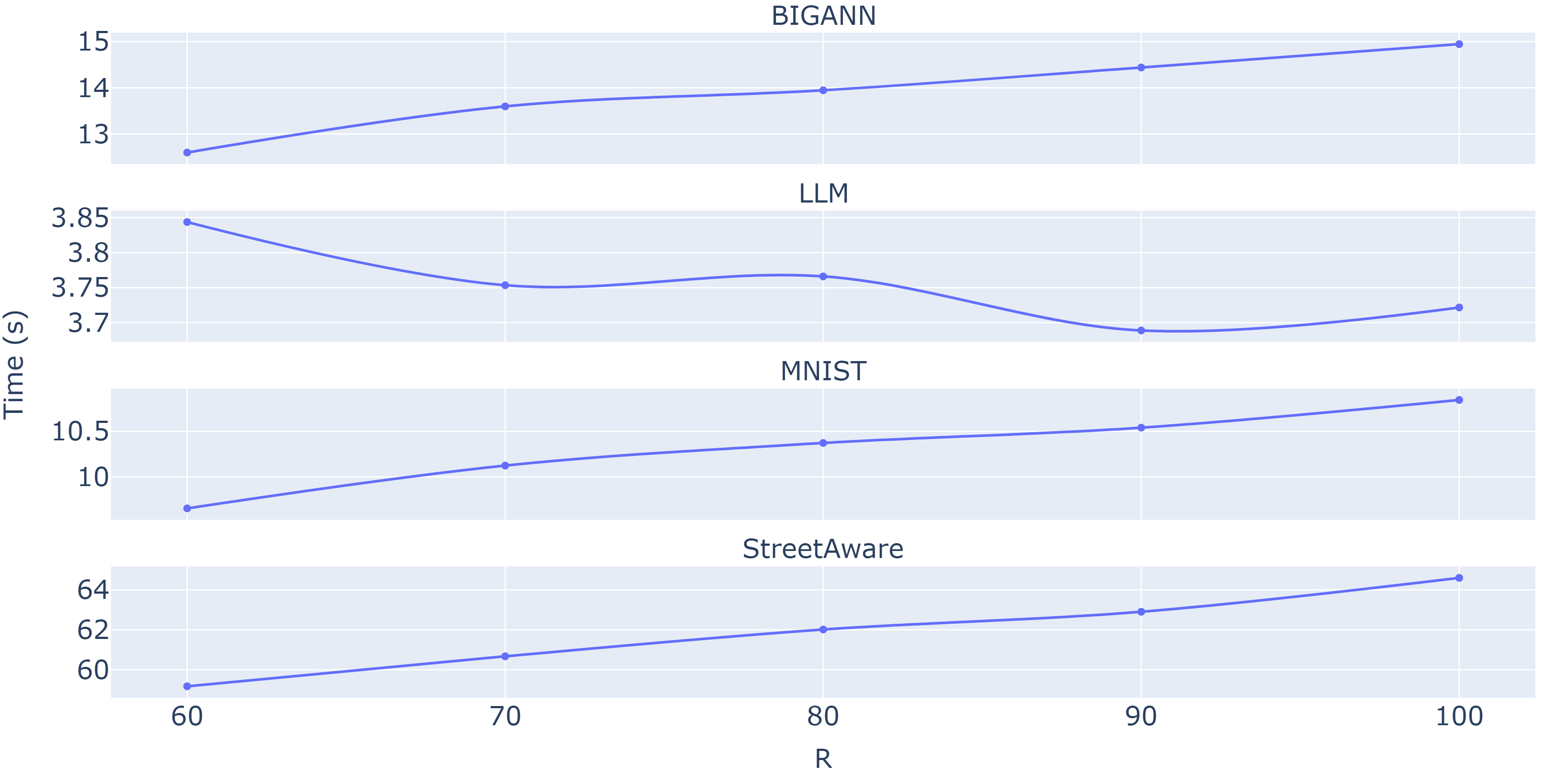}
    \caption{Time to compute the AMST with varying $R$.}
    \label{fig:impact-R-time}

    \centering
    \includegraphics[width=\linewidth]{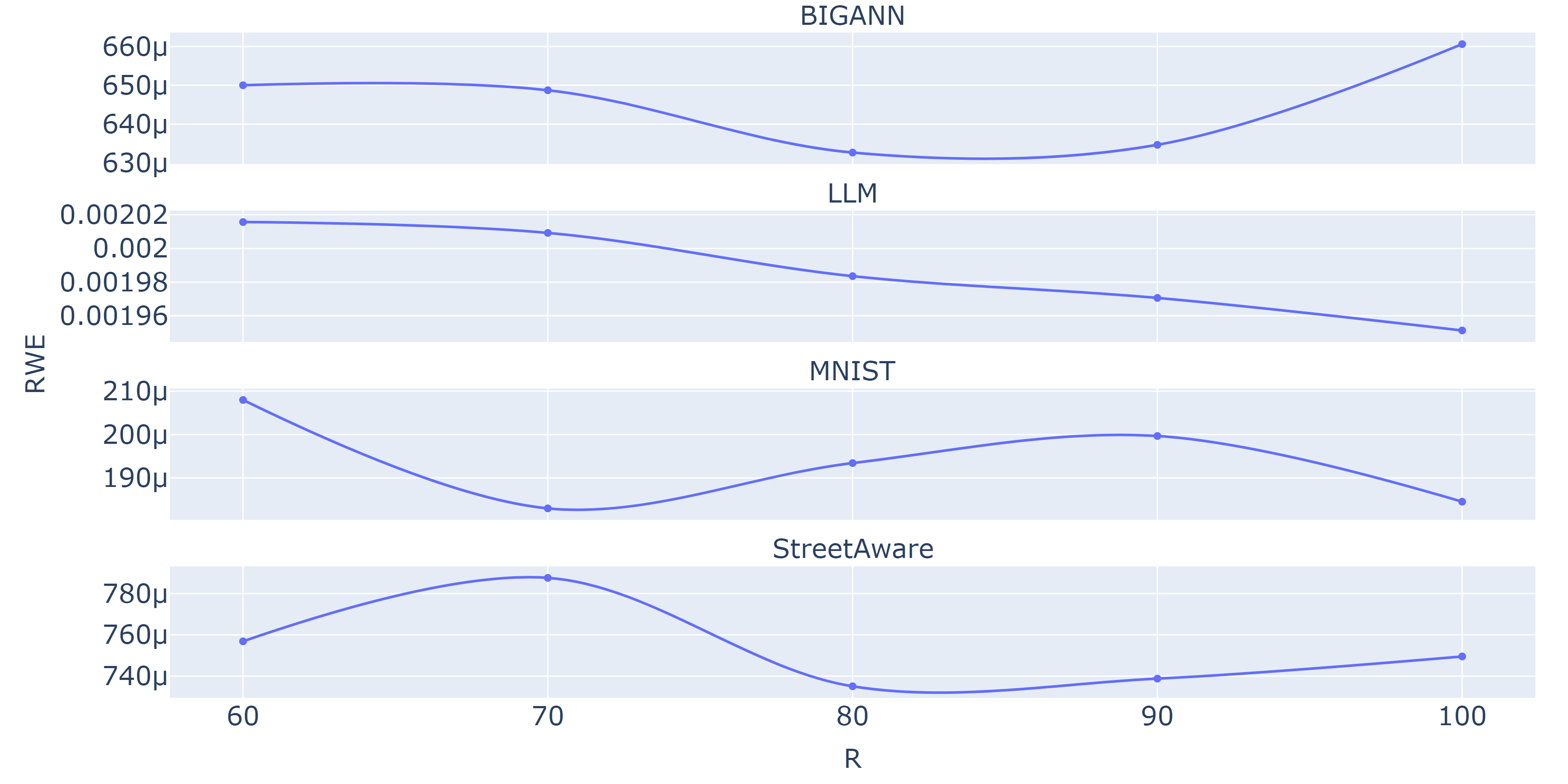}
    \caption{RWE between the AMST and EMST with varying $R$.}
    \label{fig:impact-R-RWE}

    \centering
    \includegraphics[width=\linewidth]{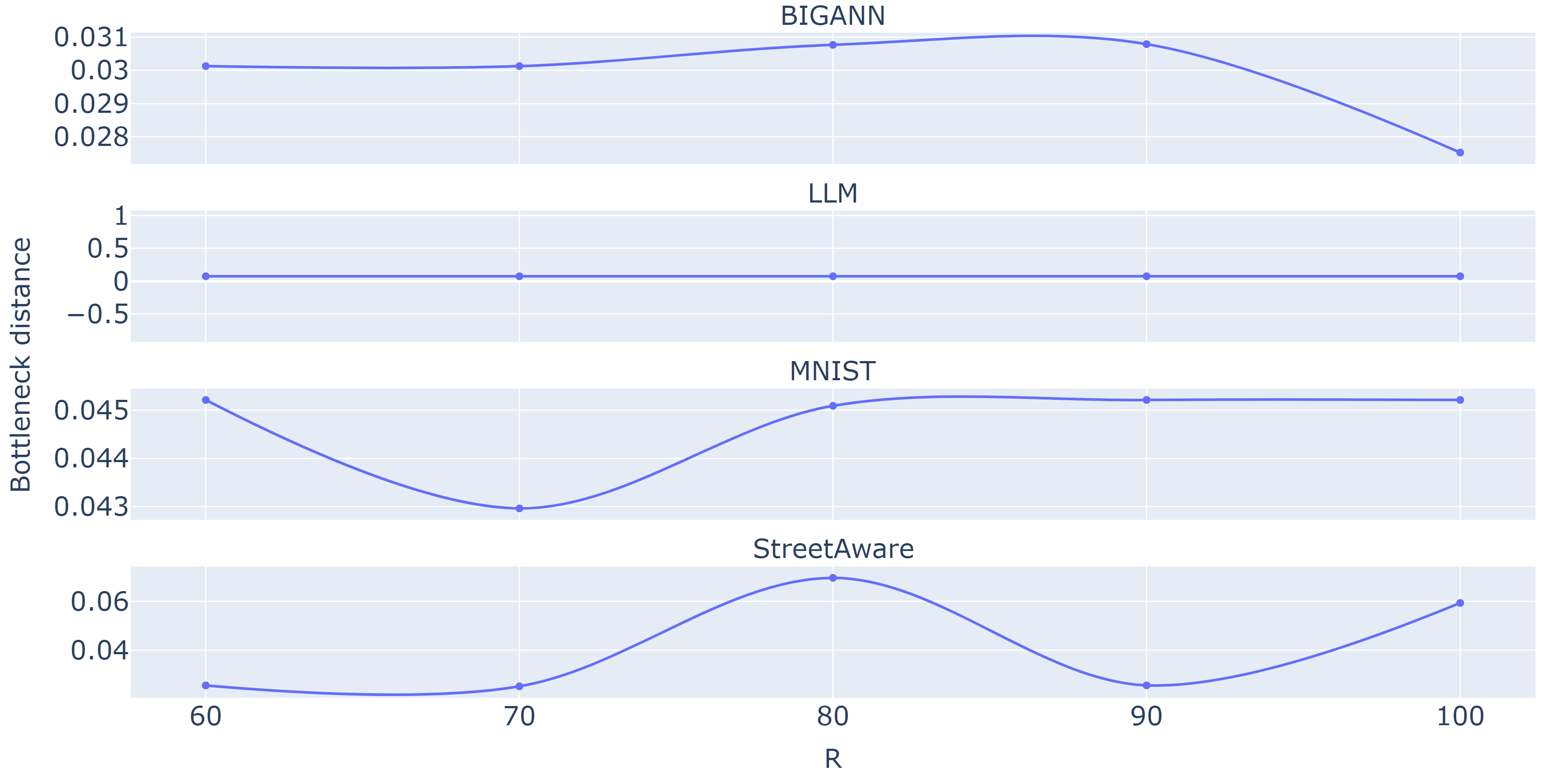}
    \caption{Bottleneck distance with varying $R$.}
    \label{fig:impact-R-bottleneck}
\end{figure}

\noindent\textbf{3.~Parameter $R$: }
This parameter defines the maximum fan out of the Vamana graph, and its suggested values are between 60 and 150, with $R$ being strictly less than $L$. Therefore for this experiment we fix $L = 100$, $\alpha=1.3$, and vary $R$ between 60 and 100.
Figures~\ref{fig:impact-R-time}, \ref{fig:impact-R-RWE}, and \ref{fig:impact-R-bottleneck} show the variation of AMST computation time, RWE, and bottleneck distance, respectively, with varying values of $R$.
As with $L$, we see that computation time increases with increasing $R$. We also see some fluctuation with respect to RWE and bottleneck distance--however, the differences in values are very small (in the order of $10^{-6}$ in case of RWE, and $10^-2$ in case of bottleneck distance) to be significant.

\subsection{Approximation Evaluation using Wasserstein Distance}

Bottleneck distance is a special case of Wasserstein distance when the $L_\infty$-norm is used. 
In the context of persistence diagrams, it captures the \textit{maximum deviation} between equivalent points in the persistence diagram. Also, since the range of this value is independent of the number of points in the input, we decided to evaluate using Bottleneck distance for the paper.
In Table~\ref{tab:wass_dist}, we look at the 1-Wasserstein distance (that uses the 1-norm) between the persistence diagrams corresponding to the different datasets. In addition to the Wasserstein distance, we also report the normalized Wasserstein distance that normalizes the metric based on the number of points in the dataset (thus allowing comparison across datasets). 
% Intuitively, this value provides the average deviation between the matches across the persistence diagrams.

\begin{table}[h]
\footnotesize
\centering
\caption{Using Wasserstein distance to evaluate the approximation quality.}
\label{tab:wass_dist}
\begin{tabular}{|l|c|c|}
\hline
\multirow{ 2}{*}{\textbf{Dataset}}     & \multicolumn{2}{c|}{\textbf{{Distance}}} \\ 
\cline{2-3}
            & \textbf{Normalized Wasserstein} & {\textbf{Wasserstein}} \\ \hline
MNIST       & 2.2$\times 10^{-2}$             & {1332.3} \\ \hline
BIGANN      & 1.9$\times 10^{-2}$             & {1762.8} \\ \hline
LLM         & 3.4$\times 10^{-3}$             & {23} \\ \hline
StreetAware & 2.0$\times 10^{-2}$             & {7442.9}  \\ \hline
\end{tabular}
\end{table}

% \balance

% \bibliographystyle{unsrt}
\bibliographystyle{abbrv-doi-hyperref}

\bibliography{references}